\documentclass[times, 10pt]{article} 
\usepackage{latex8}
\usepackage{times}

\usepackage{xspace}
\usepackage{hyperref}
\usepackage{microtype}
\usepackage{url}
\usepackage{etex}
\usepackage{amsmath}
\usepackage[english]{babel}
\usepackage{amsfonts}
\usepackage{amssymb}
\usepackage{euscript}
\usepackage{stmaryrd}
\usepackage{enumerate}
\usepackage{subfigure}
\usepackage[dvips]{graphicx}
\usepackage{QED}
\usepackage{rotating}
\usepackage{proof}
\usepackage{bussproofs}
   \EnableBpAbbreviations
   \def\extraVskip{0.3em}
     
\newcommand{\axc}[1]{\AxiomC{$#1$}}
\newcommand{\uic}[1]{\UnaryInfC{$#1$}}
\newcommand{\bic}[1]{\BinaryInfC{$#1$}}
\newcommand{\tic}[1]{\TrinaryInfC{$#1$}}

\newcommand{\kl}{\mathit{Kl}}
\newcommand{\kt}{\mathit{Kt}}

\newcommand{\ltl}{\mathit{LTL}}
\newcommand{\stl}{\mathit{STL}}
\newcommand{\mtl}{\mathit{MTL}}

\newcommand{\ctl}{\mathit{CTL}}
\newcommand{\ctlstar}{\mathit{CTL^*}}

\newcommand{\nkl}{\mathcal{N}(Kl)}
\newcommand{\nkmtl}{\mathcal{N}(K_{\mathit{MTL}})}

\newcommand{\nl}{\mathcal{N}(Kl_L)}
\newcommand{\nr}{\mathcal{N}(Kl_R)}
\newcommand{\ngen}{\mathcal{N}(Kl_G)}

\newcommand{\am}{\mathcal{M}}
\newcommand{\m}{\mathcal{M}}

\newcommand{\wld}{\mathcal{W}}

\newcommand{\val}{\mathcal{V}}
\newcommand{\prop}{\mathcal{P}}

\newcommand{\bottomE}{RAA_\bottom}
\newcommand{\emptysetE}{RAA_\emptyset}
\newcommand{\limpliesI}{\limplies\!\!I}
\newcommand{\limpliesE}{\limplies\!\!E}
\newcommand{\rimpliesI}{\rimplies\!I}
\newcommand{\rimpliesE}{\rimplies\!E}
\newcommand{\mon}{\mathit{mon}}
\newcommand{\reflequal}{\mathit{refl}\!=}
\newcommand{\irreflless}{\mathit{irrefl}\!<}
\newcommand{\transless}{\mathit{trans}\!<}

\newcommand{\bottom}{\perp}
\newcommand{\true}{\top}
\newcommand{\next}{\lhd}
\newcommand{\nextt}{N}
\newcommand{\non}{\,\sim}
\newcommand{\nmodels}{\nvDash}
\newcommand{\rimplies}{\sqsupset}
\newcommand{\limplies}{\supset}
\newcommand{\ror}{\sqcup}
\newcommand{\rand}{\sqcap}

\newcommand{\g}{\mathsf{G}}

\newcommand{\gp}{\mathsf{G}}
\newcommand{\h}{\mathsf{H}}
\newcommand{\hp}{\mathsf{H}}
\newcommand{\f}{\mathsf{F}}
\newcommand{\fp}{\mathsf{F}}
\newcommand{\p}{\mathsf{P}}
\newcommand{\pp}{\mathsf{P}}
\newcommand{\x}{\mathsf{X}}

\newcommand{\xp}{\mathsf{X}}

\newtheorem{theorem}{Theorem}
\newtheorem{definition}[theorem]{Definition}
\newtheorem{lemma}[theorem]{Lemma}

\newtheorem{proposition}[theorem]{Proposition}

\newtheorem{notation}[theorem]{Notation}
\newtheorem{remark}[theorem]{Remark}

\newcommand{\secref}[1]{Section~\ref{#1}}

\begin{document}

\title{Labeled Natural Deduction Systems for a Family of Tense Logics \\ 
(Extended Version)}

\author{Luca Vigan\`{o} \qquad Marco Volpe \\
Department of Computer Science, University of Verona, Italy \\
\{luca.vigano, marco.volpe\}@univr.it
}

\maketitle
\thispagestyle{empty}

\begin{abstract}
We give labeled natural deduction systems for a family of tense logics extending the basic linear tense logic $\kl$. We prove that our systems are sound and complete with respect to the usual Kripke semantics, and that they possess a number of useful normalization properties (in particular, derivations reduce to a normal form that enjoys a subformula property). We also discuss how to extend our systems to capture richer logics like (fragments of) $\ltl$.
\end{abstract}

\section{Introduction}\label{sec:introduction}

Hilbert-style systems, although uniform, are difficult to use in practice, especially in comparison with the more ``natural" Gentzen-style systems such as natural deduction (ND), sequent, and tableaux systems. However, devising Gentzen-style systems for modal, relevance, and other non-classical logics often requires considerable ingenuity, as well as trading uniformity for simplicity and usability. A solution to this problem is to employ \emph{labeling} techniques, which provide a general framework for presenting different logics in a uniform way in terms of Gentzen-style systems. 

The intuition is that labeling (also called prefixing, annotating or subscripting) allows one to explicitly encode additional information, of a semantic or proof-theoretical nature, that is otherwise implicit in the logic one wants to capture. So, for instance, instead of a modal formula $A$, we can consider the \emph{labeled formula (lwff)} $x:A$, which intuitively means that $A$ holds at the world denoted by $x$ within the underlying Kripke semantics. We can also use labels to specify how worlds are related in a particular Kripke model, e.g.~the \emph{relational formula (rwff)} $x < y$ states that the world $y$ is accessible from $x$.

Labeled deduction systems have been given for several non-classical logics, e.g.~\cite{AndersonBelnapDunn92,Bas+:LabelledDeduction:00,BonGor:LabelledKt:98,TableauMethods:99,Gab:LabDedSys:96,Ind:LabND4Ltl:03,Mar:PhdThesis:02,Orlowska96,Sim:PhdThesis:94,Vig:Labelled:00}, and research has focused not only on the design of systems for specific logics, but also, more generally, on the characterization of the classes of logics that can be formalized this way. General properties and limitations of labeling techniques have also been investigated. For example, \cite{Vig:Labelled:00} highlights an important trade-off between limitations and properties, which can be roughly summarized as follows. 
Assume that we have a set of rules for reasoning about the introduction and elimination of modal operators in lwffs $x:A$ such as the following rules for $\Box$, where we express $x: \Box A$ as the metalevel implication $x < y \Longrightarrow y:A$ for an arbitrary $y$ accessible from $x$ ($y$ is \emph{fresh}, i.e.~it is different from $x$ and does not occur in any assumption on which $y:A$ depends other than $x < y$):
\begin{displaymath}\footnotesize
\infer[\Box I \textrm{ (} y \textrm{ fresh)}]{x: \Box A}{\infer*{y:A}{[x < y]}} 
\qquad 
\infer[\Box E\,.]{y:A}{x: \Box A & x < y} 
\end{displaymath}
Assume also that we reason on the semantic information provided by labeling using \emph{Horn-style relational rules}
\begin{displaymath}\footnotesize
\infer{x_0 < y_0}{x_1 < y_1 & \ldots & x_n < y_n}
\end{displaymath}
where the $x_i$ and $y_i$ are labels, and $n \geq 0$ (so that the rule has no premises when $n=0$).
While restricting our systems to such Horn rules allows us to present only a subset of all possible non-classical logics, we can still capture several of the most common modal and relevance logics, and, more importantly, labeling provides an efficient general method for establishing the metatheoretical properties of these logics, including their completeness, decidability, and computational complexity. 
This method relies on the separation between the sub-system for reasoning about lwffs and the sub-system for reasoning about rwffs: derivations of lwffs can depend on derivations of rwffs (e.g.~via the $\Box$ rules), but rwffs depend only on rwffs (via the Horn rules).

In this paper, we give labeled natural deduction systems for a family of \emph{tense logics} extending the basic linear tense logic $\kl$~\cite{ResUrq:TemporalLogic:71}.  Our starting point is~\cite{Vig:Labelled:00} but it should be immediately clear that Horn rules do not suffice: even a minimal tense logic like $\kl$ requires its time points to be connected, i.e.~for any two points $x$ and $y$ either $x=y$, or $x$ is before $y$, or $y$ is before $x$. It is straightforward to see that such a property cannot be captured by a Horn rule like the one above; rather, we need non-atomic rwffs, in particular disjunction ($\ror$) of relations, and more complex rules built using a full first-order language, such as the axiom 
\begin{displaymath}\footnotesize
\infer[\mathit{conn}\,.]{\forall x.y.\, x<y \, \ror \, x=y \, \ror \, y<x}{}
\end{displaymath}
A similar situation occurs if we wish to impose irreflexivity of our worlds. And that's not all: as shown in~\cite{Vig:Labelled:00} (in the case of modal logics, but the same arguments apply here, mutatis mutandis), if we move to such a first-order language and wish to retain completeness of the resulting systems, then we need to abandon the strict separation between the sub-system for lwffs and that for rwffs (and let derivations of rwffs depend also on lwffs). As we will see in more detail below, this is best achieved by introducing a so-called \emph{universal falsum}, so that a contradiction in a world can be propagated not only to any other world but also to the relational structure to derive any rwff; and, vice versa, from a contradiction in the relational sub-system we can obtain any lwff. 

The main contributions, and the structure, of this paper are thus the following.
In \secref{sec:logic}, we give a brief presentation of the  syntax and semantics, and of a standard axiomatization, of $\kl$. In \secref{sec:system}, we give a labeled natural deduction system $\nkl$ for $\kl$, which we show to be sound and complete (extending the completeness proofs given for modal logics in~\cite{Vig:Labelled:00}). Then, in \secref{sec:normalization}, we show that $\nkl$ possesses a number of useful normalization properties; in particular, derivations reduce to a normal form that enjoys a subformula property. In \secref{sec:family}, we extend $\nkl$ to capture some interesting extensions of $\kl$, and in \secref{sec:towards-ltl} we discuss how to extend our systems to capture richer logics like (fragments of) $\ltl$. We conclude, in \secref{sec:conclusions}, by comparing with related work and discussing future work.
Detailed proofs and examples are given in an appendix.

\section{The basic linear tense logic $\kl$}\label{sec:logic}

\subsection{Syntax}

\begin{definition}
Given a set $\cal{P}$ of propositional variables, the set of \emph{well-formed $\kl$ formulas} is defined by the following Backus-Naur-form presentation, where $p \in \mathcal{P}$:
\begin{displaymath}\footnotesize
A ::= \, p \mid \bottom \mid A \limplies A \mid \g A \mid \h A\,.
\end{displaymath}
\end{definition}
Truth of a tense formula is relative to a world in a model, so, intuitively, $\g A$ holds at a world iff $A$ always holds in the future, and $\h A$ holds at a world iff $A$ always holds in the past. We will formalize this standard semantics below, but in order to give a labeled ND system for $\kl$, we extend the syntax with labels and relational symbols that capture the worlds and the accessibility relation between them. 
\begin{definition}
Let $L$ be a set of labels and let $x$ and $y$ be labels in $L$. If $A$ is a well-formed $\kl$ formula, then $x:A$ is a \emph{labeled well-formed formula} (labeled formula or lwff, for short).

The set of \emph{well-formed relational formulas} (relational formulas or rwffs, for short) is defined as follows:
\begin{displaymath}\footnotesize
\rho ::= \, x<y \mid x=y \mid \emptyset \mid \rho \rimplies \rho \mid \forall x.\, \rho\,.
\end{displaymath}
\end{definition}

We write $\varphi$ to denote a generic formula (lwff or rwff). We say that an lwff $x:A$ is \emph{atomic} when $A$ is atomic, i.e.~$A$ is a propositional variable or $A$ is $\bottom$. An rwff $\rho$ is \emph{atomic} when it does not contain any connective or quantifiers, i.e. $\rho$ is $\emptyset$ or $\rho$ has the form $x<y$ or $x=y$. 
The \emph{grade} of an lwff or rwff is the number of occurrences of connectives ($\limplies$ or $\rimplies$), operators ($\g$ or $\h$), and quantifiers ($\forall$). 
Finally, given a set of lwffs $\Gamma$ and a set of rwffs $\Delta$, we call the ordered pair $(\Gamma, \Delta)$ a \emph{proof context}.

The given syntax uses a minimal set of connectives, operators, and quantifiers. As usual, we can introduce abbreviations and use, e.g., $\non$, $\wedge$, $\vee$ and $\neg$, $\sqcap$, $\ror$, for the negation, the conjunction, and the disjunction in the labeled language and in the relational one, respectively. For instance, $\non A \equiv A\limplies \bottom$ and  $\rho' \ror \rho'' \equiv (\rho' \rimplies \emptyset) \rimplies \rho''$. We can also define $\top \equiv \non \bottom$, other quantifiers, e.g.~$\exists x.\, \rho \equiv \neg \forall x.\, \neg \rho$, and other temporal operators, e.g.~$\f A \equiv \non \g \non A$ to express that $A$ holds sometime in the future. 

\subsection{Semantics}\label{sub:kl/semantics}

\begin{definition}\label{def:kl/model}
A \emph{$\kl$ frame} is a pair $(\wld, \prec)$, where $\wld$ is a non-empty set of worlds and 
$\prec \, \subseteq \wld \times \wld$ is a binary relation that satisfies the properties of irreflexivity, transitivity and connectedness, i.e.~for all $(x, y) \in \wld^2$ we have $x=y$ or $(x,y)\in \prec$ or $(y,x)\in \prec$. 

A \emph{$\kl$ model} is a triple $(\wld, \prec, \val)$, where $(\wld, \prec)$ is a $\kl$ frame and the valuation $\val$ is a function that maps an element of $\wld$ and a propositional variable to a truth value ($0$ or $1$).
\end{definition}
In order to give a semantics for our labeled system, we need to define explicitly an interpretation of labels as worlds.
\begin{definition}\label{def:kl/truth}
Given a set of labels $L$ and a model $\m=(\wld, \prec, \val)$, an \emph{interpretation} is a function $\lambda: L \rightarrow \wld$ that maps every label in $L$ to a world in $\wld$.

Given a model $\m$ and an interpretation $\lambda$ on it, \emph{truth} for an rwff or lwff $\varphi$ is the smallest relation $\models^{\am,\lambda}$ satisfying: \\

{\footnotesize
\begin{tabular}{lcl}
$\models^{\am,\lambda} x<y$ & iff & $(\lambda(x),\lambda(y)) \in \prec$;\\
$\models^{\am,\lambda} x=y$ & iff & $\lambda(x)=\lambda(y)$;\\
$\models^{\am,\lambda} \rho_1 \rimplies \rho_2$ & iff & $\models^{\am,\lambda} \rho_1$ implies $\models^{\am,\lambda} \rho_2$;\\
$\models^{\am,\lambda} \forall x.\, \rho$ & iff & for all $y$, $\models^{\am,\lambda} \rho[y/x]$;\\
\\
$\models^{\am,\lambda} x:p$ & iff & $\val(\lambda(x),p) = 1$;\\
$\models^{\am,\lambda} x:A\limplies B$ & iff & $\models^{\am,\lambda} x:A$ implies $\models^{\am,\lambda} x:B$;\\
$\models^{\am,\lambda} x:\g A$ & iff & for all $y$, $\models^{\am,\lambda} x<y$ implies
$\models^{\am,\lambda} y:A$;\\
$\models^{\am,\lambda} x:\h A$ & iff & for all $y$, $\models^{\am,\lambda} y<x$ implies 
$\models^{\am,\lambda} y:A$.
\end{tabular}
}

Hence, $\nmodels^{\am,\lambda} x: \bot$ and $\ \nmodels^{\am,\lambda} \emptyset$. 
When $\models^{\am,\lambda} \varphi$, we say that $\varphi$ is \emph{true} in $\m$ according to the interpretation $\lambda$. By extension: \\

{\footnotesize
\begin{tabular}{lcl}
 $\models^{\am, \lambda} \Gamma$  &  iff  &   $\models^{\am,\lambda} x:A$ for all $x:A\in \Gamma$;\\
$\models^{\am, \lambda} \Delta$  &  iff  &   $\models^{\am,\lambda} \rho$ for all $\rho\in \Delta$;\\
$\models^{\am, \lambda} (\Gamma, \Delta)$  &  iff  &  $\models^{\am, \lambda} \Gamma$ and $\models^{\am, \lambda} \Delta$;\\
$\Gamma, \Delta \models^{\am, \lambda} \varphi$  &  iff  &  $\models^{\am, \lambda} (\Gamma, \Delta)$ implies $\models^{\am, \lambda} \varphi$.
\end{tabular}	
}
\end{definition}

Truth for lwffs and rwffs built using other connectives or operators can be defined in the usual manner.\footnote{Note that truth for lwffs is related to the standard truth relation for modal logics by observing that $\models^{\am} x:A$ iff $\models^{\am}_x A$.}

\subsection{An axiomatization of $\kl$}\label{subsub:axiomatization}

Several different Hilbert-style axiomatizations have been given for the logic $\kl$; the following one is taken from~\cite{ResUrq:TemporalLogic:71}:
\begin{description}\footnotesize
  \item[$\mathit{(G1)}$] $\g(A \limplies B) \limplies (\g A \limplies \g B)$
  \item[$\mathit{(G2)}$] $\non \h \non \g A \limplies A$
  \item[$\mathit{(G3)}$] $\g A \limplies \g \g A$
  \item[$\mathit{(G4)}$] $[\g (A \vee B) \; \wedge \; \g (A \vee \g B) \; \wedge \; \g (\g A \vee B)] \limplies (\g A \vee \g B)$
  \item[$\mathit{(Nec_G)}$] If $\vdash A$ then $\vdash \g A$
  \item[$\mathit{(Nec_H)}$] If $\vdash A$ then $\vdash \h A$
  \item[$\mathit{(MP)}$] If $\vdash A$ and $\vdash A \limplies B$ then $\vdash B$
\end{description}
The axiom $\mathit{(G1)}$ is standard for modal and temporal logics, while $\mathit{(G2)}$ sets the dual relation between $\g$ and $\h$, $\mathit{(G3)}$ expresses the transitivity and $\mathit{(G4)}$ the connectedness of $\g$.
For brevity, we have omitted the symmetric axioms $\mathit{(H1)}$-$\mathit{(H4)}$ that are obtained by replacing every $\g$ by $\h$ and vice versa. Moreover, every classical tautology is a tautology, and there are rules for modus ponens and necessitation for both $\g$ and $\h$.

\section{A labeled natural deduction system for $\kl$}\label{sec:system}
  
Our labeled ND system $\nkl = \nl + \nr + \ngen$ comprises of three sub-systems, whose rules are given in Figure~\ref{fig:rules-kl}.

The propositional and temporal rules of $\nl$ allow us to derive lwffs from other lwffs with the help of rwffs. The rules $\limpliesI$ and $\limpliesE$ are just the labeled version of the standard (\cite{Pra:NatDed:65,TroSch:BasicProofTheory:00}) ND rules for implication introduction and elimination, where the notion of \emph{discharged/open assumption} is also standard (e.g.~the formula $[x:A]$ is discharged in the rule $\limpliesI$). 
The rule $\bottomE$ is a labeled version of \emph{reductio ad absurdum}, where we do not enforce Prawitz's side condition that $A \neq \bot$.\footnote{See~\cite{Vig:Labelled:00} for a detailed discussion on $\bottomE$, which in particular  explains how, in order to maintain the duality of modal operators like $\Box$ and $\Diamond$, the rule must allow one to derive $x:A$ from a contradiction $\bot$ at a possibly different world $y$, and thereby discharge the assumption $x: A \limplies \bottom$.}
The temporal operators $\g$ and $\h$ share the structure of the basic introduction/elimination rules, with respect to the same accessibility relation $<$; this holds because, for instance, we express $x: \g A$ as the metalevel implication $x < y \Longrightarrow y:A$ for an arbitrary $y$ accessible from $x$ (as we did for $\Box$ in the introduction).

The relational rules of $\nr$ allow us to derive rwffs from other rwffs only. The rules $\emptysetE$, $\rimpliesI$, and $\rimpliesE$ are reductio ad absurdum and implication introduction and elimination for rwffs, while $\forall I$ and $\forall E$ are the standard rules for universal quantification, with the usual proviso for $\forall I$. There are also four axiomatic rules (or ``axioms", for short) $\reflequal$, $\irreflless$, $\transless$, and $\mathit{conn}$, which express the properties of $=$\footnote{Note that we do not need further axioms to express symmetry and transitivity of $=$, since the former can be derived by using $\mon$, $\mathit{conn}$, and $\irreflless$, and the latter by using $\mon$.} and $<$, where, for readability, we employed the  symbols for disjunction, conjunction, and negation.

The general rules of $\ngen$ allow us to derive lwffs from rwffs and vice versa.
The rule $\mon$ applies monotonicity to an lwff or rwff $\varphi$, while the rules $\mathit{uf}1$ and  $\mathit{uf}2$ export falsum (and we thus call it a \emph{universal falsum}) from the labeled sub-system to the relational one, and vice versa.\footnote{Note that the presentation of the system could be simplified by introducing a unique symbol for falsum (say $\curlywedge$), shared by the labeled and the relational sub-systems. In that case, we would not need the rules $\mathit{uf}1$ and $\mathit{uf}2$, while the rules for falsum elimination $\bottomE$ and $\emptysetE$ could be replaced by the following rule, where with $-\varphi$ we denote the negation of a generic formula (labeled or relational):
\begin{displaymath}\footnotesize
\infer[RAA_\curlywedge]{\varphi}{\infer*{\curlywedge}{[-\varphi]}}
\end{displaymath}
However, we prefer to maintain a clear separation between the two sub-systems, as it will allow us to give a simpler presentation of normalization.}			

\begin{figure*}[t]\footnotesize
  \begin{displaymath}
    \renewcommand{\arraystretch}{3}
    \begin{array}{c}
    \infer[\bottomE]{x:A}{\infer*{y:\bottom}{[x: A \limplies \bottom]}}
    \quad\
    \infer[\limpliesI]{x: A \limplies B}{\infer*{x:B}{[x: A]}}
    \quad\
    \infer[\limpliesE]{x: B}{x: A \limplies B & x: A}
    \quad\
    \infer[\g I^*]{x: \g A}{\infer*{y:A}{[x<y]}}
    \quad\
    \infer[\g E]{y:A}{x: \g A & x<y}
    \quad\
    \infer[\h I^*]{x: \h A}{\infer*{y:A}{[y<x]}}
    \quad\
    \infer[\h E]{y:A}{x: \h A & y<x} \\
    \infer[\emptysetE]{\rho}{\infer*{\emptyset}{[\rho \rimplies \emptyset]}}
    \quad\
    \infer[\rimpliesI]{\rho_1 \rimplies \rho_2}{\infer*{\rho_2}{[\rho_1]}}
    \quad\
    \infer[\rimpliesE]{\rho_2}{\rho_1 \rimplies \rho_2 & \rho_1}
    \quad\
    \infer[\forall I^*]{\forall x.\, \rho}{\rho}
    \quad\
    \infer[\forall E]{\rho[y/x]}{\forall x.\, \rho} 
    \quad\
    \infer[\reflequal]{\forall x. \, x=x}{} 
    \quad\
    \infer[\irreflless]{\forall x.\, \neg(x<x)}{}
    \\
    \infer[\transless]{\forall x.y.z.\, (x<y \rand y<z) \rimplies x<z}{}
    \quad\
    \infer[\mathit{conn}]{\forall x.y.\, x<y \, \ror \, x=y \, \ror \, y<x}{}
    \quad\
    \infer[\mon]{\varphi[y/x]}{\varphi & x=y}
    \quad\
    \infer[\mathit{uf}1]{\emptyset}{x:\bottom}
    \quad\
    \infer[\mathit{uf}2]{x:\bottom}{\emptyset}
    \end{array}
   \end{displaymath}
*In $\g I$ (respectively, $\h I$), $y$ is different from $x$ and does not occur in any assumption on which $y:A$ depends other than the discarded assumption $x<y$ (respectively, $y<x$).  \\
In $\forall I$, the variable $x$ must not occur in any open assumption on which $\rho$ depends.
  \caption{The rules of $\nkl$}\label{fig:rules-kl}
\end{figure*}   
    
\begin{definition}[Derivations and proofs]
  A \emph{derivation} of a formula (lwff or rwff) $\varphi$ from a proof context $(\Gamma,\Delta)$ in 
  $\nkl$ is a tree formed using the rules in $\nkl$, ending with $\varphi$ and depending only on 
  a finite subset of $\Gamma\cup\Delta$.  
  We then write $\Gamma, \Delta \vdash \varphi$. A derivation of $\varphi$ in $\nkl$ depending 
  on the empty set, $\vdash \varphi$, is a \emph{proof} of $\varphi$ in $\nkl$  and we then say that 
  $\varphi$ is a theorem of $\nkl$.
\end{definition}

\begin{figure*}[t]\footnotesize
  \begin{displaymath}
    \begin{array}{c}
    \infer[\f I]{x: \f A}{y:A & x<y}
    \qquad
    \infer[\f E^*]{z:B}{x:\f A & \infer*{z:B}{[y:A] [x<y]}}
    \qquad
    \infer[\p I]{x: \p A}{y:A & y<x}
    \qquad
    \infer[\p E^*]{z:B}{x:\p A & \infer*{z:B}{[y:A] [y<x]}}
    \\
    \infer[\ror I1]{\rho_1 \ror \rho_2}{\rho_1}
    \qquad
    \infer[\ror I2]{\rho_1 \ror \rho_2}{\rho_2}
    \qquad
    \infer[\ror E]{\rho}{\rho_1 \ror \rho_2 & \infer*{\rho}{[\rho_1]} & \infer*{\rho}{[\rho_2]}}
    \qquad
    \infer[\exists I]{\exists x.\, \rho}{\rho[y/x]}
    \qquad
    \infer[\exists E^*]{\rho'}{\exists x.\, \rho & \infer*{\rho'}{[\rho[y/x]]}}
    \end{array}
   \end{displaymath}
   *In $\f E$ (respectively, $\p E$), $y$ is different from $x$ and $z$, and does not occur in any assumption on which the upper occurrence of $z:B$ depends other than $y:A$ or $x<y$ (respectively, $y<x$).\\
   In $\exists E$, $y$ does not occur in any assumption on which the upper occurrence of $\rho'$ depends other than $\rho[y/x]$.
  \caption{Some derived rules}\label{fig:derived-rules}
\end{figure*}   

We will give concrete examples of derivations in the following sections. For simplicity, we will employ the rules for conjunction $\wedge$ and disjunction $\vee$, which are derived from the basic propositional rules as is standard, as well as other derived rules such as those for $\f$, $\p$, $\ror$, and $\exists$ given in Figure~\ref{fig:derived-rules}.

Since the axiomatization of $\kl$ given in \secref{subsub:axiomatization} is sound and complete, we could prove in $\nkl$ the  axioms and the rules of the axiomatization to establish the completeness of $\nkl$ indirectly (and we do so in \secref{sec:completeness-axioms}). We can, however, also  give a direct proof of the soundness and completeness of $\nkl$. In fact, by adapting standard proofs for labeled systems (see, e.g., \cite{Gab:LabDedSys:96,Sim:PhdThesis:94,Vig:Labelled:00} and the detailed proofs in the appendix, which in particular extend those for modal logics in~\cite{Vig:Labelled:00} to the case of universal falsum and other general rules that mix derivations of lwffs and rwffs), we have:
\begin{theorem}[Soundness and completeness of $\nkl$]\label{theorem:soundness-completeness}
$\nkl=\nl + \nr + \ngen$ is sound and complete, i.e.~we have that
$\, \Gamma, \Delta \vdash \varphi \;$ iff $\; \Gamma, \Delta \models^{\am, \lambda} \varphi$ for every model $\m$ and every interpretation $\lambda$.
\end{theorem}

\section{Normalization}\label{sec:normalization}

\subsection{Derivations in normal form}

We will now show that the system $\nkl$ possesses a number of useful normalization properties. To that end, we will follow the classical normalization process of~\cite{Pra:NatDed:65} as much as possible, while some adaptations are inspired by~\cite{Vig:Labelled:00}. We begin by simplifying the proofs by restricting the applications of some of the rules.

\begin{lemma}\label{lem:kl/restrictions}
If $\Gamma, \Delta \vdash \varphi$, then there exists a derivation of $\varphi$ from $(\Gamma, \Delta)$ where: $(i)$ the conclusions of applications of $\bottomE$, $\emptysetE$, and $\mon$ are atomic; $(ii)$ $\mon$ is not applied to lwffs of the form $x:\bottom$.
\end{lemma}
The system obtained from $\nkl$ by restricting the rules $\bottomE$, $\emptysetE$, and $\mon$ according to this lemma is equivalent to $\nkl$.
From now on, we will thus consider only this restricted system and keep calling it $\nkl$.

\begin{figure*}[t]\footnotesize
$$  
  \begin{array}{c}
   
    \begin{array}{c}

    \begin{array}{ccc}
    \def\defaultHypSeparation{\hskip .00000001in}	    
    \axc{\scriptstyle [x<y]^1}
			\noLine
	  \uic{\scriptstyle \pi}
      \noLine
    \uic{\scriptstyle y:A}
			\RightLabel{$\scriptstyle \gp I^1$}
		\uic{\scriptstyle x:\gp A}
	  \axc{\scriptstyle \pi_2}
			\noLine
	 	\uic{\scriptstyle x<z}
			\RightLabel{$\scriptstyle \gp E$}		
	 	\bic{\scriptstyle z:A}
	 \DisplayProof
	    &
	  \rightsquigarrow
	    &
    \def\defaultHypSeparation{\hskip .00000001in}		    
	  \axc{\scriptstyle \pi_2}
			\alwaysNoLine
	 	\uic{\scriptstyle x<z}
	 	\uic{\scriptstyle \pi[z/y]}
	 	\uic{\scriptstyle z:A}
	 \DisplayProof
  \end{array}
  \\
  \\
  \footnotesize
  $(a) Reduction for the detour\ $ {\scriptstyle \gp I} / {\scriptstyle \gp E}
  \end{array}
  
\begin{array}{c}

  \begin{array}{ccc}
  \def\defaultHypSeparation{\hskip .00000001in}	
	  \axc{\scriptstyle \pi}
	   \noLine	  	  
	  \uic{\scriptstyle x:\bottom}
			\RightLabel{$\scriptstyle \bottomE$}	  
	  \uic{\scriptstyle y:\bottom}
			\RightLabel{$\scriptstyle \mathit{uf}1$}
	 	\uic{\scriptstyle \emptyset}
	 \DisplayProof	 
	    &
	  \rightsquigarrow
	    &
	  \def\defaultHypSeparation{\hskip .00000001in}	
	  \axc{\scriptstyle \pi}
	   \noLine	  	    
	  \uic{\scriptstyle x:\bottom}
			\RightLabel{$\scriptstyle \mathit{uf}1$}
	 	\uic{\scriptstyle \emptyset}
	 \DisplayProof	 
	  \end{array}
  \\
  \\ 
  \\  
  \footnotesize
  $(b) A reduction for falsum-rules$
  \end{array}
 
 \begin{array}{c}
  \begin{array}{ccc}
\def\defaultHypSeparation{\hskip .00000001in} 
	  \axc{\scriptstyle \pi_1}
	   \noLine	  	  
	  \uic{\scriptstyle \varphi}
	  \axc{\scriptstyle \pi_2}
	   \noLine	  	  	  
	  \uic{\scriptstyle x=y}
			\RightLabel{$\scriptstyle \mon$}
	  \bic{\scriptstyle \varphi[y/x]}
	  \axc{\scriptstyle \pi_3}
	   \noLine	  	  	  
	  \uic{\scriptstyle y=z}
			\RightLabel{$\scriptstyle \mon$}
	 	\bic{\scriptstyle \varphi[z/x]}
	 \DisplayProof
	 
	    &
	  \rightsquigarrow
	    &
	    
\def\defaultHypSeparation{\hskip .00000001in}	    
	  \axc{\scriptstyle \pi_1}
	   \noLine	  	  
	  \uic{\scriptstyle \varphi}
	  \axc{\scriptstyle \pi_2}
	   \noLine	  	  	  
	  \uic{\scriptstyle x=y}
	  \axc{\scriptstyle \pi_3}
	   \noLine	  	  	  
	  \uic{\scriptstyle y=z}
			\RightLabel{$\scriptstyle \mon$}
	  \bic{\scriptstyle x=z}
			\RightLabel{$\scriptstyle \mon$}
	 	\bic{\scriptstyle \varphi[z/x]}
	 \DisplayProof		 
	\end{array}  
	
  \\
  \\
  \\
  \footnotesize
  $(c) Reduction for the rule $\mon
  \end{array}	
	
  \end{array}
$$
  \caption{Examples of reductions}\label{fig:reductions}
\end{figure*}

The ND systems given in~$\cite{Vig:Labelled:00}$ for families of modal and relevance logics are based on a strict separation between the labeled and the relational sub-systems (i.e.~derivations of lwffs can depend on derivations of rwffs, but not vice versa). This separation is possible thanks to the restriction to relational theories that are Horn theories. Our system $\nkl$ does not allow for such a separation, since the rules for universal falsum let relational derivations depend also on labeled ones. Thus, more complex derivations are possible, which implies that with respect to~\cite{Vig:Labelled:00} we need to consider more forms of detours and hence more forms of reductions. 

\begin{definition}
We say that a formula $\varphi$ is a \emph{maximal formula} in a derivation when it is both the conclusion of an introduction rule and the major premise of an elimination rule.
  
  We define the notion of \emph{label position} for labels occurring in a formula $\varphi$ to which the rule $\mon$ is applied. By the restrictions of Lemma~\ref{lem:kl/restrictions}, $\varphi$ can have the form $(i)\,x:p$, $(ii)\,x<y$, or $(iii)\,x=y$. We say that $x$ has label position $1$ in $(i)$, $(ii)$ and $(iii)$, and $y$ has label position $2$ in $(ii)$ and $(iii)$.
  
  A derivation is in \emph{pre-normal form} (is a \emph{pre-normal derivation}) if it has no maximal formulas and in every sequence of $\mon$ applications, all the applications which concern variables with the same label position occur consecutively.
\end{definition}
The notion of pre-normal derivation embodies the elimination of standard detours (given by a couple of introduction/elimination rule applications on the same connective or operator) and an ordering of $\mon$ applications that aims at eliminating $\mon$ detours, i.e. two or more applications of $\mon$ which concern variables with the same label position. Note that, since $\mon$ is only applied to atomic formulas of the form described above, once we have eliminated maximal formulas, the case of a sequence of $\mon$ applications is the only case in which we can have this kind of detour.
\begin{lemma}\label{lem:pre-normal}
  Every derivation in $\nkl$ reduces to a derivation in pre-normal form.
\end{lemma}
\begin{proof} (Sketch)
 First, we iteratively apply proper reductions (an example is in Figure~\ref{fig:reductions}$(a)$) that remove maximal formulas. Then the lemma follows by observing that applications of $\mon$ in a sequence can be permuted as shown in Figure~\ref{fig:mon-ordering} in the appendix.
\end{proof}
\begin{definition}
We call \emph{falsum-rules} the rules $\bottomE$, $\emptysetE$, $\mathit{uf}1$, and $\mathit{uf}2$.
We say that a formula $\varphi$ is a \emph{redundant formula} in a derivation when: $(i)$ $\varphi$ is both the conclusion and the premise of a falsum-rule; or $(ii)$ $\varphi$ is both the conclusion and the major premise of a $\mon$ carrying out two substitutions in the same label position (see Figure~\ref{fig:reductions}$(c)$).

A derivation is in \emph{normal form} (is a \emph{normal derivation}) iff it is in pre-normal form and does not contain any redundant formula.
\end{definition}

\begin{theorem}\label{th:kl/normal-form}
Every derivation in $\nkl$ reduces to a derivation in normal form.
\end{theorem}
\begin{proof} (Sketch)
By Lemma~\ref{lem:pre-normal}, every derivation reduces to a pre-normal derivation. Then we can apply permutative reductions (examples in Figure~\ref{fig:reductions}$(b)-(c)$) that remove redundant formulas. More details are given in the appendix.
\end{proof}

Normal derivations in $\nkl$ have a well-defined structure that has a number of desirable properties. In particular, there is an ordering on the application of the rules, which we can exploit to prove a subformula property for our system. To that end, we adapt the standard definitions of subformula and track as follows:
\begin{definition}
$B$ is a \emph{subformula} of $A$ iff (i) $A$ is $B$; (ii) $A$ is $A_1 \limplies A_2$ and $B$ is a subformula of $A_1$ or $A_2$; (iii) $A$ is $\g A_1$ and $B$ is a subformula of $A_1$; or (iv) $A$ is $\h A_1$ and $B$ is a subformula of $A_1$. We say that $y:B$ is a \emph{subformula} of $x:A$ iff $B$ is a subformula of $A$.
  
$\rho_2$ is a \emph{subformula} of $\rho_1$ iff (i) $\rho_1$ is $\rho_2$; (ii) $\rho_1$ is $\rho_1' \rimplies \rho_1''$ and $\rho_2$ is a subformula of $\rho_1'$ or $\rho_1''$; or (iii) $\rho_1$ is $\forall x.\, \rho$ and $\rho_2$ is a subformula of $\rho$.

Given a derivation $\pi$ in $\nkl$, a \emph{track} in $\pi$ is a sequence of formulas $\varphi_1, \ldots, \varphi_n$ such that:

(i) $\varphi_1$ is an assumption of $\pi$, an axiom, or the conclusion of a universal falsum rule ($\mathit{uf}1$ or $\mathit{uf}2$);

(ii) $\varphi_i$ stands immediately above $\varphi_{i+1}$ and is the major (or the only) premise of a rule for $1 \leq i < n$;

(iii) $\varphi_n$ is the conclusion of $\pi$, the premise of a universal falsum rule, or the minor premise of a rule.

We call a track $\varphi_1, \ldots, \varphi_n$ a \emph{labeled track} when each $\varphi_i$ is an lwff and a \emph{relational track} when each $\varphi_i$ is an rwff.
\end{definition}

\begin{figure}
\begin{center}
  \includegraphics[scale=0.5, angle=0]{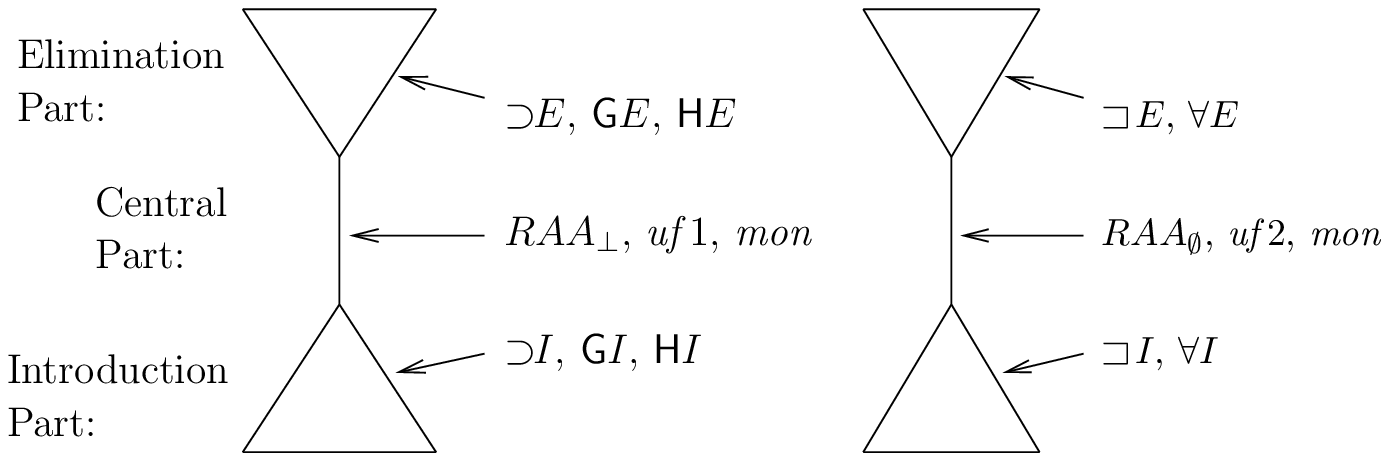}
\caption{The structure of a labeled track (left) and that of a relational track (right)
\label{fig:kl/clessidre}}
\end{center}
\end{figure}

In other words, a track can only pass through the major premises of rules and it ends at the first minor premise of a rule, or at an application of universal falsum, or at the conclusion of $\pi$.
The following lemmas formalize properties of the structure of the tracks and specify the way in which the tracks are linked one to each other.
\begin{lemma}\label{lem:kl/tracks-properties}
Let $\pi$ be a normal derivation, and let $t$ be a track $\varphi_1, \ldots, \varphi_n$ in $\pi$. Then $t$ consists of three (possibly empty) parts: (1) an \emph{elimination part}, (2) a \emph{central part}, and (3) an \emph{introduction part} (see Figure~\ref{fig:kl/clessidre}) where:

(i) each $\varphi_i$ in the elimination part is the major premise of an elimination rule and contains $\varphi_{i+1}$ as a subformula;

(ii) each $\varphi_j$ in the introduction part except the last one is the premise of an introduction rule and is a subformula of $\varphi_{j+1}$;

(iii) each $\varphi_k$ in the central part is atomic and is the premise of a falsum-rule or the major premise of a $\mon$;

(iv) the central part contains at most one application of falsum-rules;

(v) tracks originating from an application of $\mathit{uf}1$ or $\mathit{uf}2$ have an empty elimination part;

(vi)  tracks ending in an application of $\mathit{uf}1$ or $\mathit{uf}2$ have an empty introduction part.
\end{lemma}

\begin{figure}
\begin{center}
  \begin{tabular}{cccc}
  \includegraphics[scale=0.5, angle=0]{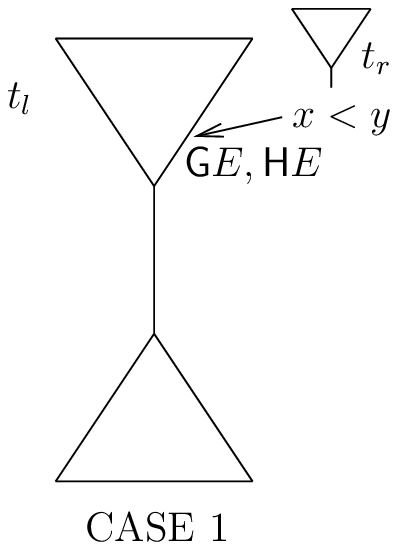}
  &
  \includegraphics[scale=0.5, angle=0]{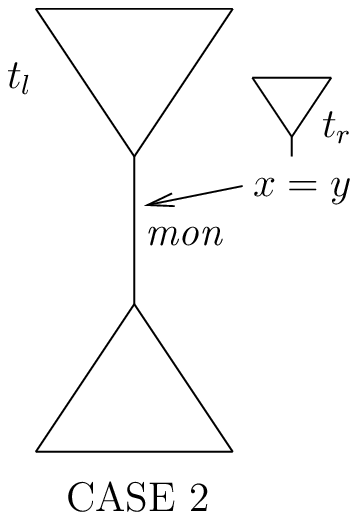}
  &
  \includegraphics[scale=0.5, angle=0]{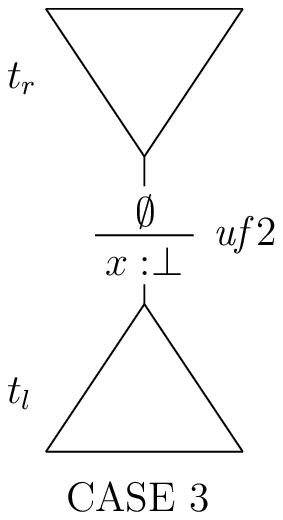}
  &
  \includegraphics[scale=0.5, angle=0]{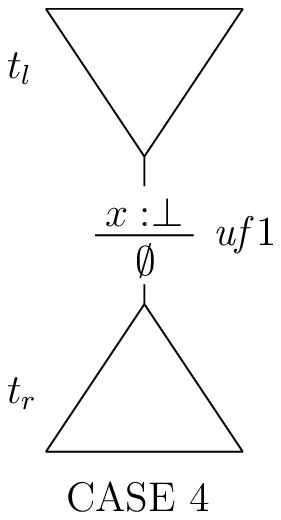}
  \end{tabular}
\caption{Possible connections between labeled tracks $t_l$ and relational tracks $t_r$
\label{fig:kl/track-links}}
\end{center}
\end{figure}

\begin{lemma}\label{lem:kl/tracks-linking}
Let $t_l$ be a labeled track and $t_r$ a relational track in a derivation $\pi$. Then $t_l$ and $t_r$ can be connected in one of the following ways (shown in Figure~\ref{fig:kl/track-links}):

(i) the last formula in $t_r$ is the minor premise of a $\g E$ or of a $\h E$ whose major premise is a formula in the elimination part of $t_l$;

(ii) the last formula in $t_r$ is the minor premise of a $\mon$ whose major premise is a formula in the central part of $t_l$;

(iii) $t_r$ ends with an application of $\mathit{uf}2$ and the conclusion of that application is the first formula in $t_l$;

(iv) $t_l$ ends with an application of $\mathit{uf}1$ and the conclusion of that application is the first formula in $t_r$. 
\end{lemma}
\begin{proof}
The statement follows trivially by observing that $\g E$, $\h E$, $\mon$, $\mathit{uf}1$, and $\mathit{uf}2$ are the only rules that mix labeled and relational formulas and that, by Lemma~\ref{lem:kl/tracks-properties}, such rules can be applied only in a specific part of a track.
\end{proof}

\subsection{The subformula property}\label{sub:kl/subformula-property}

To prove a subformula property for $\nkl$, we adapt further standard definitions: 
\begin{definition}\label{def:kl/subformula-property}
Given a derivation $\pi$ in $\nkl$, the \emph{main thread} is the sequence $t_1, \ldots, t_n$ of tracks such that: (1) the first formula in $t_1$ is an assumption or an axiom; (2) $t_i$ and $t_{i+1}$ are connected by means of an application of $\mathit{uf}1$ or $\mathit{uf}2$, for $1 \leq i \leq (n-1)$; and (3) the last formula in $t_n$ is the conclusion of $\pi$.

Let $\pi$ be a derivation of $\varphi$ from $(\Gamma, \Delta)$ in $\nkl$, $S_L$ be the set of subformulas of the formulas in $\Gamma$ (or in $\Gamma \cup \{\varphi\}$ if $\varphi$ is a labeled formula), and $S_R$ be the set of subformulas of the formulas in $\Delta \cup Ax$ (or in $\Delta \cup Ax \cup \{\varphi\}$ if $\varphi$ is a relational formula), where $Ax$ is the set of axioms used in $\pi$. We say that $\pi$ enjoys the \emph{subformula property} iff
\begin{enumerate}\itemsep=-3pt
\item for all lwffs $y:B$ used in the derivation $\pi$: \\
(i) $B \in S_L$; or \\
(ii) $B$ is an assumption $D \limplies \bottom$ discharged by an application of $\bottomE$ where $D \in S_L$; or \\
(iii) $B$ is an occurrence of $\bottom$ obtained by $\limpliesE$ from an assumption $D \limplies \bottom$ discharged by an application of $\bottomE$, where $D \in S_L$; or \\
(iv) $B$ is an occurrence of $\bottom$ obtained by an application of $\bottomE$ that does not discharge any assumption; or \\
(v) $B$ is an occurrence of $\bottom$ obtained by an application of $\mathit{uf}2$;
\item for all rwffs $\rho$ used in the derivation $\pi$: \\
(i) $\rho \in S_R$; or \\
(ii) $\rho$ is an assumption $\rho_1 \rimplies \bottom$ discharged by an application of $\emptysetE$ where $\rho_1 \in S_R$; or \\
(iii) $\rho$ is an occurrence of $\emptyset$ obtained by $\rimpliesE$ from an assumption $\rho' \rimplies \emptyset$ discharged by an application of $\emptysetE$, where $\rho' \in S_R$; or \\
(iv) $\rho$ is an occurrence of $\emptyset$ obtained by an application of $\mathit{uf}1$; or \\
(v) $\rho$ is obtained by an application of $\mon$.
\end{enumerate}
\end{definition}

\begin{lemma}\label{lem:kl/subformula-property}
Every normal derivation in $\nkl$ satisfies the subformula property.
\end{lemma}
\begin{proof} 
This follows immediately from the standard proof~\cite{Pra:NatDed:65}, which is based on the introduction of an ordering of the tracks in a normal derivation depending on their distance from a main thread. In our case, a main thread contains not only labeled formulas and we have to consider more cases than in the standard proof, given that the central part of a track can have a more complex structure (as it can also contain applications of $\mathit{uf}1$, $\mathit{uf}2$, and $\mon$).
\end{proof}

This lemma shows that although normal derivations in $\nkl$ have a more complex structure than normal derivations in ND systems for classical logic~\cite{Pra:NatDed:65} and ND systems for families of modal and relevance logics~\cite{Vig:Labelled:00}, they have still a well-defined structure and satisfy a subformula property. It is important to remark that the special cases added to the definition of subformula property (i.e.~formulas can be derived by applications of $\mathit{uf}1$, $\mathit{uf}2$, or $\mon$) do not compromise automatic proof search completely, given that such cases can occur only in a limited section of a normal derivation (i.e.~the central part of a track).
  
We also note that the presence of axioms (and in particular the fact that they are expressed in a full first-order language) makes our proof of normalization more complex and our results weaker. Thus, it is not possible to use it as a means to show the consistency of the system or the validity of an interpolation theorem, as can be done for systems in~\cite{Vig:Labelled:00}, where relational properties are expressed by Horn rules and we have only atomic axioms.

\section{A family of tense logics}\label{sec:family}

The basic linear tense logic $\kl$ leaves unanswered many fundamental natural questions about the structure of time. However, the labeling framework allows us to express several further relational properties in a straightforward and clean way, i.e.~by only adding the corresponding relational axioms to the relational sub-system. In particular, we will now show how to extend $\nkl$ to capture the extensions of $\kl$ with: 
a first/final point;
unbounded time;
dense time; and
discrete time (where we adopted the classification of~\cite{ResUrq:TemporalLogic:71}).\footnote{It is worth to mention that in~\cite{BonGor:LabelledKt:98}, Bonnette and Gor\'e give a labeled sequent system for the minimal tense logic $\kt$ that can easily capture any combination of the reflexive, transitive, euclidean, symmetric and serial extensions of the logic. We have not considered all of these properties of the accessibility relation here, but the missing ones can be added straightforwardly thanks to the modularity of our system, which we exploit to capture the extensions towards $\ltl$ we consider in the remainder of the paper. The labeling discipline of~\cite{BonGor:LabelledKt:98} is different from ours and is tailored to a lean Prolog implementation of their sequent systems. In contrast, we focus here on the proof-theoretical aspects of our ND systems and leave an implementation for future work.} 
  
\paragraph{$\kl$ with a first/final point}\label{sub:kl-first-point}

The semantics of $\kl$ is given by means of temporal structures where nothing is said about the existence of a first or a final point. To express the existence of such points, we can add the following axioms\footnote{The existence of a first (or a final) point is often expressed by adding a constant to the language. For example, we could introduce a constant $0$ for the first point and an axiom stating that $\forall y. \, \neg (y<0)$.
We prefer not to modify the language and keep the treatment of this property closer to that of other ones.} to the relational sub-systems:
\begin{displaymath}\footnotesize
\infer[\mathit{first}]{\exists x.\, \forall y.\, \neg (y<x)}{}
\qquad 
\infer[\mathit{final}\,.]{\exists x.\, \forall y.\, \neg (x<y)}{}
\end{displaymath}
The two axioms do not affect each other; thus we can decide to add both or just one of them to the system, according to the logic we want to represent.

Soundness of the extended systems is straightforward, since the axioms mirror the properties that the models of the extended logic are required to satisfy. To show completeness, it suffices to extend the canonical model construction presented for $\nkl$ (see~\secref{sub:appendix-completeness})
to consider also the new relational axioms. Alternatively, we can simply prove completeness by proving the corresponding (see, e.g., \cite{Ven:TemporalLogic:01}) Hilbert-style axioms \emph{(having a first point)} and \emph{(having a final point)} that are given in Figure~\ref{fig:axioms}. In Figure~\ref{fig:first-point}, we show the derivation for the first one (the proofs of the two axioms are symmetric).
Also the normalization procedure of \secref{sec:normalization} can still be applied to the extended system: we have just to consider the possibility of more relational axioms.

\begin{figure}\footnotesize
\begin{center}
\begin{tabular}{ll}
\emph{(having a first point)}
&
$\h \bottom \, \vee \; \p \h \bottom$ \\
\emph{(having a final point)}
&
$\g \bottom \, \vee \; \f \g \bottom$ \\
\emph{(left-seriality)}
& $\p \true$
\\
\emph{(right-seriality)}
& $\f \true$
\end{tabular}
\begin{tabular}{ll}
\emph{(left-density)}
& $\p A \limplies \p \p A$
\\
\emph{(right-density)}
&
$\f A \limplies \f \f A$
\\
\emph{(left-discreteness)}
&
$(\p \true \wedge A \wedge \g A) \limplies (\p \g A)$
\\
\emph{(right-discreteness)}
&
$(\f \true \wedge A \wedge \h A) \limplies (\f \h A)$
\end{tabular}
\end{center}
\caption{Some axioms for extensions of $\kl$}
\label{fig:axioms}
\end{figure}

\begin{figure*}\footnotesize
\begin{center}
	  \def\defaultHypSeparation{\hskip .2in}
	  
	  \axc{}
	   \RightLabel{$\scriptstyle \mathit{first}$}
	 	\uic{\scriptstyle \exists x.\forall y.\, \neg (y<x)}
	 	
	  \axc{}
	   \RightLabel{$\scriptstyle \mathit{conn}$}
	 	\uic{\scriptstyle \forall x.y.\, x<y \ror x=y \ror y<x}
	   \RightLabel{$\scriptstyle \forall E$}
	 	\uic{\scriptstyle \forall y.\, t<y \ror t=y \ror t<x}
	   \RightLabel{$\scriptstyle \forall E$}
	 	\uic{\scriptstyle t<s \ror t=s \ror s<t}

	 	\axc{\scriptstyle [\forall y.\, \neg (y<s)]^2}
	   \RightLabel{$\scriptstyle \forall E$}
	 	\uic{\scriptstyle \neg (t<s)}
	 	\axc{\scriptstyle [t<s]^3}
	   \RightLabel{$\scriptstyle \neg E$}
	 	\bic{\scriptstyle \emptyset}

	 	\axc{\scriptstyle [t=s \ror s<t]^3}

	 	\axc{\scriptstyle \pi_1}
	 	 \noLine
	 	\uic{\scriptstyle \emptyset}

	 	\axc{\scriptstyle \pi_2}
      \noLine
	 	\uic{\scriptstyle \emptyset}

	   \RightLabel{$\scriptstyle \ror E^4$}
	 	\tic{\scriptstyle \emptyset}
	 	
	   \RightLabel{$\scriptstyle \ror E^3$}
	 	\tic{\scriptstyle \emptyset}
	 	
	   \RightLabel{$\scriptstyle \exists E^2$}
	 	\bic{\scriptstyle \emptyset}
	 	
	   \RightLabel{$\scriptstyle \mathit{uf}2$}
	 	\uic{\scriptstyle t:\bottom}
	 	
	   \RightLabel{$\scriptstyle {\bottomE}^1$}
	 	\uic{\scriptstyle t:\hp \bottom \vee \pp \hp \bottom}
\DisplayProof
\end{center}

\begin{tabular}{ll}

    where $\pi_1$ is:
    &
    and $\pi_2$ is:
\\
	  \def\defaultHypSeparation{\hskip .05in}
    
	 	\axc{\scriptstyle [t:\pp \true \wedge \hp \pp \true]^1}
	   \RightLabel{$\scriptstyle \wedge E$}
	 	\uic{\scriptstyle \neg (t:\pp \true)}
    \axc{\scriptstyle [t=s]^4}
	   \RightLabel{$\scriptstyle \mon$}
	 	\bic{\scriptstyle (s:\pp \true)}

	 	\axc{\scriptstyle [\forall y.\, \neg (y<s)]^2}
	   \RightLabel{$\scriptstyle \forall E$}
	 	\uic{\scriptstyle \neg (q<s)}
    \axc{\scriptstyle [q<s]^5}
	   \RightLabel{$\scriptstyle \neg E$}
	 	\bic{\scriptstyle \emptyset}
	   \RightLabel{$\scriptstyle \mathit{uf}2$}
	 	\uic{\scriptstyle s:\bottom}
	   \RightLabel{$\scriptstyle \pp E^5$}
	 	\bic{\scriptstyle s:\bottom}
	   \RightLabel{$\scriptstyle \mathit{uf}1$}
	 	\uic{\scriptstyle \emptyset}
	 	\DisplayProof      
&
	  \def\defaultHypSeparation{\hskip .05in}

	 	\axc{\scriptstyle [t:\pp \true \wedge \hp \pp \true]^1}
	   \RightLabel{$\scriptstyle \wedge E$}
	 	\uic{\scriptstyle t: \hp \pp \true}
	 	\axc{\scriptstyle [s<t]^4}
	   \RightLabel{$\scriptstyle \hp E$}
	 	\bic{\scriptstyle s: \pp \true}

	 	\axc{\scriptstyle [\forall y.\, \neg (y<s)]^2}
	   \RightLabel{$\scriptstyle \forall E$}
	 	\uic{\scriptstyle \neg(r<s)}
	 	\axc{\scriptstyle [r<s]^6}
	   \RightLabel{$\scriptstyle \neg E$}
	 	\bic{\scriptstyle \emptyset}
	   \RightLabel{$\scriptstyle \mathit{uf}2$}
	 	\uic{\scriptstyle s:\bottom}
	   \RightLabel{$\scriptstyle \pp E^6$}
	 	\bic{\scriptstyle s:\bottom}
	   \RightLabel{$\scriptstyle \mathit{uf}1$}
	 	\uic{\scriptstyle \emptyset}
	 	\DisplayProof
\end{tabular}
\caption{Derivation of the modal axiom for first point}
\label{fig:first-point}
\end{figure*}

\paragraph{$\kl$ with unbounded time}\label{sub:kl-infinite-time}
Conversely, we can express the fact that the sequence of time points is unbounded, towards the past and/or towards the future. This corresponds to adding the conditions of seriality on the left and/or on the right, i.e.~every point has a predecessor and/or a successor. For this, we can add two relational axioms corresponding to the axioms for left and right seriality given in Figure~\ref{fig:axioms}:
\begin{displaymath}\footnotesize
\infer[\mathit{lser}]{\forall x.\exists y.\, y<x}{}
\qquad
\infer[\mathit{rser}\,.]{\forall x.\exists y.\, x<y}{}
\end{displaymath}
As an example, we show completeness for \emph{(right-seriality)}, where $\pi$ is some proof of $s:\true$ based on a proof of $\true$ or $A \vee\! \non A$ in classical logic (see, e.g., \cite{Pra:NatDed:65, TroSch:BasicProofTheory:00}):
\begin{displaymath}\footnotesize
         \begin{array}{c}

	  \def\defaultHypSeparation{\hskip .2in}

				  \axc{}
  					\RightLabel{$\scriptstyle \mathit{rser}$}
					\uic{\forall x. \exists y.\, x<y}
  					\RightLabel{$\scriptstyle \forall E$}
					\uic{\exists y.\, t<y}					
					
					\axc{\pi}
            \noLine
					\uic{s: \true}
					
					\axc{[t<s]^1}
  					\RightLabel{$\scriptstyle \fp I$}
  				\bic{t:\f \true}
					
  					\RightLabel{$\scriptstyle \exists E^1$}
  				\bic{t:\f \true}
					
					\DisplayProof 					
				\end{array}			
\end{displaymath}
                        
\paragraph{$\kl$ with dense time}\label{sub:dense-kl}
Another constraint that we can impose on relational structures is that the flow of time is dense, i.e.~between any two points we can find a third point:
\begin{displaymath}\footnotesize
\infer[\mathit{dens}\,.]{\forall x.y.\, x<y \rimplies \exists z.\, x<z \rand z<y}{}
\end{displaymath}
Figure~\ref{fig:axiom-rdensity} in the appendix shows the proof of the axiom for \emph{(right-density)}; the proof for \emph{(left-density)} can be obtained in a symmetric way by using the same axiom ($\mathit{dens}$).

\paragraph{$\kl$ with discrete time}\label{sub:discrete-kl}
Finally, we can express discreteness both towards the past and towards the future:
\begin{displaymath}\footnotesize
\infer[\mathit{ldiscr}]{\forall x.y.\, x<y \rimplies \exists z. \, z<y \rand \neg \exists u.\, (z<u \rand u<y)}{}
\end{displaymath}
\begin{displaymath}\footnotesize
\infer[\mathit{rdiscr}\,.]{\forall x.y.\, x<y \rimplies \exists z. \, x<z \rand \neg \exists u.\, (x<u \rand u<z)}{}
\end{displaymath}
We omit the proof of completeness for the corresponding axioms.

\section{Towards $\ltl$}\label{sec:towards-ltl}

We have seen that ND systems for several extensions of $\kl$ can be given by extending the ``base system"  $\nkl$. This is not the case for all the possible extensions, however, as some properties, e.g.~continuity or finite intervals, are second-order properties~\cite{Ven:TemporalLogic:01}
and thus require an appropriate higher-order relational language. We now briefly discuss whether (and how) it is possible to extend $\nkl$ to capture a richer logic like (fragments of) $\ltl$.

\paragraph{$\mtl$: a subset of $\ltl$}
For brevity, we restrict our attention to future temporal operators only (but the extension to the past is straightforward) and begin by considering the system $\nkl$ extended with the axioms $\mathit{rdiscr}$ and $\mathit{rser}$ so that the flow of time is discrete and unbounded towards the future (in this case, the presence of $\mathit{rser}$ allows us to simplify $\mathit{rdiscr}$ to $\forall x.\, \exists z. \, x<z \rand \neg \exists u.\, (x<u \rand u<z)$). We can express in our syntax the relation \emph{next} in terms of the relation $<$ (see, e.g.,~\cite{Gor:Temporal:00}), i.e.~we can introduce a relational symbol $\next$ (with the meaning of \emph{immediately precedes}) as an abbreviation:
\begin{displaymath}\footnotesize
  s \next t \equiv s<t \, \rand \, \forall x.\, \neg(s<x) \ror \neg(x<t)\,.
\end{displaymath}
This allows us to enrich the language with an operator $\x$, whose semantics can be given without having to introduce a specific relation for it in the definition of a model. We just need to require that models for this logic are $\kl$ models where $<$ is also discrete and serial on the right, and extend the definition of truth with:
\begin{displaymath}\footnotesize
\models^{\am,\lambda} x: \x A \quad \textrm{iff} \quad \models^{\am,\lambda} x \next y\ \textrm{ and } \models^{\am,\lambda} y: A\,.
\end{displaymath}
Rules for introduction and elimination of $\x$ can now be given in a clean way, with the usual freshness proviso for $\xp I$:\footnote{The fact that every time point has one (and only one) immediate successor follows from right-discreteness, right-seriality, and connectedness, and it allows one to express rules for $\x$ both in a universal and in an existential formulation. We give here the universal one.}

\begin{displaymath}\footnotesize
\infer[\xp I \textrm{ (} y \textrm{ fresh)}]{x: \x A}{\infer*{y:A}{[x\next y]}}
\qquad
\infer[\xp E\,.]{y:A}{x: \x A & x\next y}
\end{displaymath}
The logic that we capture in this extended system, which we call $\nkmtl$, is not $\ltl$ yet. We are able to express the existence of an immediate successor, but we miss a way to say that between any two points (related by $\prec$) there can be only a finite sequence of points related one to each other by the relation \emph{next}. We would need to express the finite interval property, but this is a second-order property, as observed above.

In~\cite{Mar:PhdThesis:02}, a subset of $\ltl$ called \emph{Small Temporal Logic}, or $\stl$ for short, is introduced and given a natural deduction system. The reasons behind the definition of $\stl$ are the difficulties arising from dealing with the induction principle (relating $\next$ and $<$) that is needed in order to represent $\ltl$. While the semantics of $\ltl$ can be given by considering Kripke structures defined over a relation of successor (denoted by $\nextt$) and by defining $\prec$ as the least transitive closure of $\nextt$, in the semantics of $\stl$ the relation $\prec$ is just required to contain $\nextt$. It follows that a rule for induction is not needed in a system for $\stl$.

It is easy to verify that $\nkmtl$ is complete with respect to the semantics of $\stl$. Moreover, it can be proven to correspond to a logic ``larger'' than $\stl$ for which the condition of linearity (or connectedness) on the relation $\prec$ holds: we call this logic \emph{Medium Temporal Logic} $\mtl$.\footnote{An axiomatization of $\mtl$ can be obtained, as shown in~\cite{Gor:Temporal:00}, by adding the following axioms to those given for future-time $\kl$:
\begin{description}\itemsep=-1pt\footnotesize
  \item[$\mathit{(K_{\xp})}$] $\qquad \x(A \limplies B) \limplies (\x A \limplies \x B)$
  \item[$\mathit{(FUNC)}$] $\,(\x \non A \limplies \non \x A) \wedge (\non \x A \limplies \x \non A)$
  \item[$\mathit{(REC_{\gp})}$] $\,\;(\g A \limplies \x (A \wedge \g A)) \wedge (\x (A \wedge \g A) \limplies \g A)$
\end{description}
}
We could also introduce rules for the operators \emph{since} and \emph{until}, but they would be quite complex and problematic from a proof-theoretical point of view; see~\cite{BCRV08} for a labeled tableaux system for a distributed temporal logic that comprises full $\ltl$, and~\cite{Bol+:AutomatedLTL:07} for tableaux-like ND rules for $\ltl$.

\paragraph{$\ltl$}
Several systems of labeled natural deduction for $\ltl$, e.g.~\cite{Bol+:LTL:06,Bol+:AutomatedLTL:07,Mar:PhdThesis:02}, introduce an induction rule like the following
\begin{displaymath}\footnotesize
\infer[\mathit{ind}]{y:A}{x:A & x<y & \infer*{x'':A}{[x<x'][x'\next x''][x':A]}}
\end{displaymath}
which does not operate at a purely relational level.
Some remarks are worth about a solution like this. First of all, the rule $\mathit{ind}$ adds some more points of contact between the labeled and the relational sub-systems and leads to a failure of normalization. Moreover, one can show that the axiom of connectedness is not needed anymore since it is in a way ``contained'' in the induction principle. In fact, the axiom~$(3)$ 
\begin{displaymath}\footnotesize
  \non \g (\g A \limplies B) \limplies \g (\g B \limplies A)
\end{displaymath}
of \emph{weak connectedness} must obviously hold in $\ltl$, for it can be subsumed by the induction axiom (see, e.g., \cite{Gol:LogicsTimeComp:87}). Thus, in the case we want to use a rule like $\mathit{ind}$ to capture $\ltl$, it seems more reasonable to follow a different approach that avoids both the extension of the relational language to a first-order language and the introduction of the universal falsum. In other words, we could have a system for $\ltl$ which uses only Horn rules in the relational theory (from which it follows that we have only atomic rwffs and no relational falsum) but extends the labeled sub-systems with a rule for induction that mixes labeled and relational premises.

\section{Conclusions}\label{sec:conclusions}

We have already discussed some works that are related to the labeled ND systems for tense logics that we have given here (for which, summarizing, we have proved not only soundness and completeness, but also a number of useful proof-theoretical properties, and for which we also discussed extensions leading up to $\ltl$).
As we observed, the main difficulties in applying the labeled deduction framework in the context of linear temporal logics arise from the need of expressing the condition of \emph{connectedness} in the case of the basic linear tense logic $\kl$ (see~\cite{Ind:LabND4Ltl:03} for a discussion) and the \emph{induction} principle in the case of $\ltl$. In fact, \cite{Ind:LabND4Ltl:03} gives a fairly complex labeled tableaux system for the logic $\kl$ (called there the linear temporal logic $Kt4.3$), which is analytical in that it only comprises elimination rules for temporal operators and can be used as a decision procedure. In contrast, the main distinctive feature of our approach is the extension of a fixed base system for the temporal operators with relational rules that express the relational properties of the considered logic. This, in particular, allows for uniform and modular proofs of meta-theoretic properties for families of logics, like the proofs we have given here. Moreover, it makes our systems amenable to extensions to other logics as we have begun investigating towards $\ltl$ and to the branching-time logics $\ctl$ and $\ctlstar$. To that end, we plan to capitalize on the labeled ND systems for $\ltl$ given in~\cite{Bol+:LTL:06, Mar:PhdThesis:02}, which both make use of a specific rule for induction.

\newpage

\appendix
\section{Proofs}

In this appendix we give the full proofs of the lemmas and theorems given in the body of the paper.
In Sections~\ref{sub:appendix-soundness} and~\ref{sub:appendix-completeness}, we give the proofs for the soundness and completeness of the system $\nkl$ (Theorem \ref{theorem:soundness-completeness}) and for the completeness of the extensions of $\nkl$ (\secref{sec:family}). In~\secref{sub:appendix-normalization}, we give proofs for the normalization results presented in~\secref{sec:normalization}.

	\subsection{Soundness}
	 \label{sub:appendix-soundness}
\begin{theorem}
	\label{th:kl/nkl-soundness}
	$\nkl = \nl + \nr + \ngen$ is sound, i.e. it holds:
	\begin{enumerate}[(i)]
	 \item $\Gamma, \Delta \vdash \rho \;$ implies $\Gamma, \Delta \models^{\am, \lambda} \rho \,$ for every model $\m$ and every interpretation $\lambda$;
	 \item $\Gamma, \Delta \vdash x:A \;$ implies $\Gamma, \Delta \models^{\am, \lambda} x:A \,$ for every model $\m$ and every interpretation $\lambda$.
	\end{enumerate}  
\end{theorem}
\begin{proof}
	\begin{enumerate}[(i)]
	 \item The proof is by induction on the structure of the derivation of $\rho$. The base case is when $\rho \in \Delta$ and is trivial. There is one step case for every axiom or rule. The axioms $\mathit{conn}$, $\transless$, and $\irreflless$ directly refer to the properties of connectedness, transitivity, and irreflexivity of $\kl$ models (Definition \ref{def:kl/model}) and thus are trivially sound, while $\reflequal$ and $\mon$ preserve soundness by definition of $\models^{\am, \lambda}x=y$ (Definition \ref{def:kl/truth}).
	 
	 Consider the case of an application of $\emptysetE$
	 $$
	  \def\defaultHypSeparation{\hskip .2in}	 
	 					\alwaysNoLine
					\axc{\scriptstyle \Gamma \; \Delta \; [\rho \rimplies \emptyset]^1}
					\uic{\scriptstyle \pi}
					\uic{\scriptstyle \emptyset}
					\alwaysSingleLine
						\RightLabel{$\scriptstyle \emptysetE^1$}
					\uic{\scriptstyle \rho}
					\DisplayProof
   $$
where $\Delta_1 = \Delta \cup \{ \rho \rimplies \emptyset \}$. By the induction hypothesis, $\Gamma, \Delta_1 \models^{\am, \lambda} \emptyset$ for every model $\m$ and every interpretation $\lambda$. Let us consider an arbitrary model $\m$ and an arbitrary interpretation $\lambda$; we assume $\models^{\am, \lambda} (\Gamma, \Delta)$ and prove $\models^{\am, \lambda} \rho$. Since $\nvDash^{\am, \lambda} \emptyset$, from the induction hypothesis we obtain $\nvDash^{\am, \lambda} (\Gamma,\Delta_1)$, that, given the assumption $\models^{\am, \lambda} (\Gamma, \Delta)$, leads to $\nvDash^{\am, \lambda} \rho \rimplies \emptyset$, i.e. $\models^{\am, \lambda}\rho$ and $\nvDash^{\am, \lambda}\emptyset$ by Definition \ref{def:kl/truth}. 
	 
	 The cases for  $\rimpliesI$, $\rimpliesE$, $\forall I$ and $\forall E$ follow by simple adaptations of the standard proofs for classical logic.	 
	 
	 Finally, consider the case of an application of $\mathit{uf}1$
	 $$
	        \axc{\scriptstyle \Gamma \; \Delta}
	        \alwaysNoLine
					\uic{\scriptstyle \pi}
					\uic{\scriptstyle x:\bottom}
					\alwaysSingleLine
						\RightLabel{$\scriptstyle \mathit{uf}1$}
					\uic{\scriptstyle \emptyset}
					\DisplayProof
   $$
   for a proof context $(\Gamma, \Delta)$ and some label $x$. By the induction hypothesis, we have $\Gamma, \Delta \models^{\am, \lambda} x:\bottom$ for every $\m$ and every $\lambda$. Given a generic model $\m$ and a generic interpretation $\lambda$, we can write $\nmodels^{\am, \lambda} x:\bottom$; it follows that $\nmodels^{\am, \lambda}(\Gamma, \Delta)$ and then also $\Gamma, \Delta \models^{\am, \lambda} \emptyset$ by Definition \ref{def:kl/truth}.
	 \item As in (i), by induction on the structure of the derivation of $x:A$. The base case is trivial and there is a step case for every rule of the labeled system. The cases of introduction and elimination of connectives and that of universal falsum are as in (i).

	 Consider an application of the rule $\g I$
	 $$	     
	        \axc{\scriptstyle \Gamma \; \Delta \; [x<y]^1}
	        \alwaysNoLine
					\uic{\scriptstyle \pi}
					\uic{\scriptstyle y:A}
					\alwaysSingleLine
						\RightLabel{$\scriptstyle \gp I^1$}
					\uic{\scriptstyle x:\gp A}
					\DisplayProof
   $$
   where  $\Gamma,\Delta_1  \vdash y:A$ with $y$ fresh and with $\Delta_1 = \Delta \cup \{x < y\}$.  By the induction hypothesis, for every model $\m$ and every interpretation $\lambda$ it holds $\Gamma, \Delta \models^{\am, \lambda} y:A$. We let $\lambda$ be any interpretation such that $\models^{\am, \lambda}(\Gamma,\Delta)$ and show that $\models^{\am, \lambda} x: \g A$. Let $w$ be any world such that $\lambda(x) \prec w$.  Since $\lambda$ can be trivially extended to another interpretation (still called $\lambda$ for simplicity) by setting $\lambda(y)=w$, the induction hypothesis yields $\models^{\am, \lambda} y:A$, and thus $\models^{\am, \lambda} x: \g A$.

	 
	 Finally, consider an application of the rule $\g E$
	 $$
          \axc{\scriptstyle \Gamma_1 \; \Delta_1}
          \alwaysNoLine
	        \uic{\scriptstyle \pi_1}	       
					\uic{\scriptstyle x:\gp A}
					\axc{\scriptstyle \Gamma_2 \; \Delta_2}
					\uic{\scriptstyle \pi_2}
					\uic{\scriptstyle x<y}
					\alwaysSingleLine
						\RightLabel{$\scriptstyle \gp E$\,.}
					\bic{\scriptstyle y:A}
					\DisplayProof
   $$
	Let $\m$ be an arbitrary model and $\lambda$ an arbitrary interpretation. If we assume $\models^{\am, \lambda}(\Gamma_1 \cup \Gamma_2, \Delta_1 \cup \Delta_2)$, then from the induction hypotheses we obtain $\models^{\am, \lambda}x:\g A$ and $\models^{\am, \lambda}x<y$, and thus $\models^{\am, \lambda}y:A$ by Definition \ref{def:kl/truth}.
	
	The treatment of $\h I$ and $\h E$ is analogous.
	\end{enumerate}
\end{proof}

  \subsection{Completeness}
    \label{sub:appendix-completeness}
In the following, in order to simplify the derivations, we will use some derived rules. We show here, as an example, how to derive the rules $\f I$ and $\f E$ (see Figure~\ref{fig:derived-rules}) from the rules for introduction/elimination of $\g$. We remind that the following equivalence holds: $\f A \, \equiv \, \non \g \non A \, \equiv \, (\g (A \limplies \bottom)) \limplies \bottom$.

The rule
	 $$
          \axc{\scriptstyle y:A}
          \axc{\scriptstyle x<y}
						\RightLabel{$\scriptstyle \fp I$}
	        \bic{\scriptstyle x: \f A}
	        \DisplayProof
   $$
can be derived as follows
	 $$
          \axc{\scriptstyle [x: \g (A \limplies \bottom)]^1}
          \axc{\scriptstyle x<y}
						\RightLabel{$\scriptstyle \gp E$}
	        \bic{\scriptstyle y:A \limplies \bottom}	       
	        \axc{\scriptstyle y:A}
	          \RightLabel{$\scriptstyle \limplies E$}
					\bic{\scriptstyle y:\bottom}
	          \RightLabel{$\scriptstyle \bottomE$}
					\uic{\scriptstyle x:\bottom}					
	          \RightLabel{$\scriptstyle \limplies I^1$}
					\uic{\scriptstyle x:\g (A\limplies \bottom)\limplies \bottom}
					\DisplayProof \qquad \,
   $$
   while an application of $\f E$
	 $$
          \axc{\scriptstyle x:\f A}
          \axc{\scriptstyle [y:A]\,[x<y]}
            \noLine
          \uic{\pi}
            \noLine
          \uic{\scriptstyle z:B}
						\RightLabel{$\scriptstyle \fp E$}          
          \bic{\scriptstyle z:B}
	        \DisplayProof
   $$
   can be replaced by the following derivation
	 $$
          \axc{\scriptstyle x:\g (A \limplies \bottom) \limplies \bottom}
          \axc{\scriptstyle [z:B \limplies \bottom]^1}
          \axc{\scriptstyle [y:A]^3\,[x<y]^2}
            \noLine
          \uic{\pi}
            \noLine
          \uic{\scriptstyle z:B}
	          \RightLabel{$\scriptstyle \limplies E$}
					\bic{\scriptstyle z:\bottom}
	          \RightLabel{$\scriptstyle \bottomE$}
					\uic{\scriptstyle y:\bottom}
	          \RightLabel{$\scriptstyle \limplies I^3$}
					\uic{\scriptstyle y: A \limplies \bottom}
	          \RightLabel{$\scriptstyle \g I^2$}
					\uic{\scriptstyle x: \g(A \limplies \bottom)}
	          \RightLabel{$\scriptstyle \limplies E$}
					\bic{\scriptstyle x: \bottom}
	          \RightLabel{$\scriptstyle {\bottomE}^1$}
					\uic{\scriptstyle z: B}					
	        \DisplayProof
   $$
   
\subsubsection{Completeness by canonical model construction}

In the following, slightly abusing notation, we will write $\varphi \in (\Gamma,\Delta)$ whenever $\varphi \in \Gamma$ or $\varphi \in \Delta$, and write $x \in (\Gamma,\Delta)$ whenever the label $x$ occurs in some $\varphi \in (\Gamma, \Delta)$.

\begin{definition}
  \label{def:kl/consistent-pc}
  A proof context $(\Gamma, \Delta)$ is \emph{$\nkl$-consistent} iff $\Gamma, \Delta \nvdash x:\bottom$ for every $x$, and it is \emph{$\nkl$-inconsistent} otherwise.
\end{definition}
Note that we can have inconsistency also by deriving $\emptyset$ in the relational system; given the rules $\mathit{uf}1$ and $\mathit{uf}2$ for universal falsum, also this case is captured by the previous definition.

For simplicity, in the following we will omit the ``$\nkl$'' and simply speak of consistent and inconsistent proof contexts.
\begin{proposition}
	\label{prop:kl/phi-or-not-phi}
	Let $(\Gamma, \Delta)$ be a consistent proof context. Then:
	\begin{enumerate}[(i)]
	 \item for every $x$ and every $A$, either $(\Gamma \cup \{ x:A \}, \Delta)$ is consistent or $(\Gamma \cup \{ x:\non A \}, \Delta)$ is consistent;
	 \item for every relational formula $\rho$ , either $(\Gamma, \Delta \cup \{\rho\})$ is consistent or $(\Gamma, \Delta \cup \{\neg \rho\})$ is consistent.
	\end{enumerate}
\end{proposition}
\begin{proof}
	\begin{enumerate}[(i)]
	 \item Suppose that both $(\Gamma \cup \{ x:A \}, \Delta)$ and $(\Gamma \cup \{ x:\non A \}, \Delta)$ are inconsistent. Then from $\Gamma \cup \{ x:A \}, \Delta \vdash x:\bottom$, by applying the rule $\limpliesI$, we get $\Gamma, \Delta \vdash x:\non A$. Similarly, from $\Gamma \cup \{ x:\non A \}, \Delta \vdash x:\bottom$, by applying the rule $\bottomE$, we get $\Gamma, \Delta \vdash x:A$.\\
	 But, if both $x:A$ and $x:\non A$ are derivable in the proof context $(\Gamma, \Delta)$, then it also holds $\Gamma, \Delta \vdash x:\bottom$, by $\sim E$. It follows that the original proof context $(\Gamma, \Delta)$ had to be inconsistent (contradiction).
	 \item The proof for the relational case is analogous and is obtained by using the corresponding relational rules i.e. $\rimpliesI$, $\emptysetE$ and $\neg E$.
	\end{enumerate}
\end{proof}

\begin{definition}
  \label{def:kl/max-consistent-pc}
  A proof context $(\Gamma, \Delta)$ is \emph{maximally consistent} iff the following three conditions hold:
  \begin{enumerate}
   \item $(\Gamma, \Delta)$ is consistent,
   \item for every relational formula $\rho$, either $\rho \in \Delta$ or $\neg\rho \in \Delta$,
   \item for every $x$ and every $A$, either $x:A \in \Gamma$ or $x: \non A \in \Gamma$.
  \end{enumerate}
\end{definition}

Completeness follows by a Henkin--style proof, where a canonical model
\begin{displaymath}
\m^C = (\wld^C, \prec^C, \val^C)
\end{displaymath}
is built from a proof context $(\Gamma,\Delta)$ to show that $(\Gamma,\Delta) \nvdash \varphi$ implies 
$\Gamma, \Delta \nmodels^{\am^C,\lambda^C} \varphi$ for every formula $\varphi$.

In standard proofs for unlabeled modal, temporal, and for other non-classical logics, the set $\wld^C$ is obtained by progressively building maximally consistent sets of formulas, where consistency is locally checked within each set.  In our case, given the presence of lwffs and rwffs, we modify the Lindenbaum lemma to extend $(\Gamma,\Delta)$ to one single maximally consistent context 
$(\Gamma^*, \Delta^*)$, where consistency is ``globally'' checked also against the additional
assumptions in $\Delta $.\footnote{We consider only consistent proof contexts.  If $(\Gamma, \Delta)$ is 
inconsistent, then $\Gamma, \Delta \vdash \varphi$ for all $\varphi$, and thus completeness immediately holds for all lwffs and rwffs.} The elements of $\wld^C$ are then built by partitioning $\Gamma^*$ and $\Delta^*$ with respect to the labels, and the relation $\prec^C$ between the worlds is defined by exploiting the information in $\Delta^*$.  

In the Lindenbaum lemma for predicate logic, a maximally consistent and $\omega$-complete 
set of formulas is inductively built by adding for every formula $\non \forall x.\, A$ a 
\emph{witness} to its truth, namely a formula $\non A[c/x]$ for some new individual constant 
$c$. This ensures that the resulting set is $\omega$-complete, i.e.~that if, for every closed 
term $t$, $A[t/x]$ is contained in the set, then so is $\forall x.\, A$. A similar procedure applies 
here not only for rwffs $\neg \forall x.\, \rho$, but also in the case of lwffs of the form $x: \non \g A$.  That is, together with $x: \non \g A$ we consistently add $y: \non A$ and $x < y$ for some new $y$, which acts as a \emph{witness world} to the truth of $x: \non \g A$.  This ensures that the maximally
consistent context $(\Gamma^*, \Delta^*)$ is such that if $x < z \in (\Gamma^*, \Delta^*)$ implies 
$z:B \in (\Gamma^*, \Delta^*)$ for every $z$, then $x: \g B \in (\Gamma^*, \Delta^*)$, as shown in
Lemma~\ref{lem:kl/properties-max-cons-pc} below.  Note that in the standard completeness proof for unlabeled modal logics, for instance, one instead considers a canonical model $\am^C$ and shows that if $\mathcal{W}_1 \in \wld^C$ and $\am^C,\mathcal{W}_1 \vDash \non \g A$, then $\wld^C$ also contains a world $\mathcal{W}_2$ accessible from $\mathcal{W}_1$ that serves as a witness world to the truth of $\non \g A$ at $\mathcal{W}_1$, i.e.~$\am^C,\mathcal{W}_2 \vDash \non A$.

\begin{lemma}
  \label{lem:kl/max-consistent}
	Every consistent proof context $(\Gamma, \Delta)$ can be extended to a maximally consistent proof context $(\Gamma^*, \Delta^*)$.
\end{lemma}
\begin{proof}
We first 
extend the language of $\nkl$ with infinitely many new constants for witness terms and for witness worlds. Let $t$ range over the original terms, $s$ range over the new constants for witness terms, and $r$ range over both; further, let $w$ range over labels, $v$ range over the new constants for witness worlds, and $u$ range over both. All these may be subscripted. 
Let $\varphi_1$, $\varphi_2$, ... be an enumeration of all lwffs and rwffs in the extended language; when $\varphi_i$ is $u:A$, we write $\non \varphi_i$ for $u: \non A$.

We iteratively build a sequence of consistent proof contexts by defining $(\Gamma_0, \Delta_0)$ = $(\Gamma, \Delta)$ and $(\Gamma_{i+1}, \Delta_{i+1})$ to be:
\begin{itemize}
 \item $(\Gamma_i, \Delta_i)$, if $(\Gamma_i \cup \{\varphi_{i+1}\}, \Delta_i)$ is inconsistent; else
 \item $(\Gamma_i \cup \{ u: \non \g A, v: \non A \}, \Delta_i \cup \{ u < v \})$ for a $v$ not occurring in $(\Gamma_i \cup \{ u: \non \g A \}, \Delta_i)$ if $\varphi_{i+1}$ is $u: \non \g A$; else
 \item $(\Gamma_i \cup \{ u: \non \h A, v: \non A \}, \Delta_i \cup \{ v < u \})$ for a $v$ not occurring in $(\Gamma_i \cup \{ u: \non \h A \}, \Delta_i)$ if $\varphi_{i+1}$ is $u: \non \h A$; else
 \item $(\Gamma_i, \Delta_i \cup \{ \neg \forall x.\, \rho, \neg \rho[s/x] \})$ for an $s$ not occurring in $(\Gamma_i, \Delta_i \cup \{ \neg \forall x.\, \rho \})$ if $\varphi_{i+1}$ is $\non \forall x.\, \rho$; else
 \item $(\Gamma_i \cup \{\varphi_{i+1}\}, \Delta_i)$ if $\varphi_{i+1}$ is an lwff or $(\Gamma_i, \Delta_i \cup \{\varphi_{i+1}\})$ if $\varphi_{i+1}$ is an rwff.
\end{itemize}
Now define
$$
(\Gamma^*,\Delta^*) = (\bigcup_{i\geq 0}\Gamma_i,\bigcup_{i\geq 0}\Delta_i)\,.
$$
We show that the proof context $(\Gamma^*, \Delta^*)$ is maximally consistent, i.e. it verifies the three conditions of Definition \ref{def:kl/max-consistent-pc}.
	\begin{enumerate}[(i)]
	 \item First we prove that our construction preserves consistency by showing that every $(\Gamma_i, \Delta_i)$ is consistent. The only interesting cases are when $\varphi_{i+1}$ is one of $\non \g A$, $\non \h A$, or $\neg \forall x.\, \rho$. We only consider the first case, since the second one is symmetrical, and the third is very similar.
	 
	 If $(\Gamma_i \cup \{ u:\non \g A \}, \Delta_i)$ is consistent, then so is $(\Gamma_i \cup \{ u: \non \g A, v: \non A \})$ for a $v$ not occurring in $(\Gamma_i \cup \{ u: \non \g A\}, \Delta_i)$. By contraposition, suppose that
	 $$
	 	\Gamma_i \cup \{ u:\non \g A, v:\non A \}\, , \, \Delta_i \cup \{ u<v \} \, \vdash \, u_j:\bottom 
	 $$
	 by a derivation $\pi$ (where $v$ does not occur in $(\Gamma_i \cup \{ u: \non \g A\}, \Delta_i)$).
	 Then in $\nkl$ we can have a derivation like the following:
$$
	  \def\defaultHypSeparation{\hskip .00001in}
	 	\alwaysNoLine
    \axc{\scriptstyle \Gamma_i \quad \Delta_i \quad u:\non \gp A \quad [v:\non A]^1 \quad [u<v]^2}
		\uic{\scriptstyle  \pi}
		\uic{\scriptstyle  u_j:\bottom}
		\alwaysSingleLine
			\RightLabel{$\scriptstyle {\bottomE}^1$}
		\uic{\scriptstyle  v:A}
			\RightLabel{$\scriptstyle \gp I^2$}
		\uic{\scriptstyle  u:\gp A}
		\axc{\scriptstyle  u:\non \gp A}
			\RightLabel{$\scriptstyle \non E$}
		\bic{\scriptstyle  u:\bottom}
\DisplayProof
$$    
   This shows that $(\Gamma_i \cup \{ u:\non \g A \}, \Delta_i)$ is inconsistent, which is not the case.
   \item Consider an rwff $\rho$. Suppose that both $\rho \notin \Delta^*$ and $\neg\rho \notin \Delta^*$ hold. Let $\rho$ be $\varphi_{i+1}$ for some $i$ in our enumeration of formulas and $\neg\rho$ be $\varphi_{j+1}$. Now suppose $i<j$ (the other case is symmetric). $\rho \notin \Delta^*$ implies that $(\Gamma_i, \Delta_i \cup \{\varphi_{i+1}\})$ is inconsistent. Given that in our inductive construction we only add formulas to the proof context, i.e.~$\Delta_i \subseteq \Delta_j$, we have that $(\Gamma_j, \Delta_j \cup \{\varphi_{i+1}\})$ is also inconsistent. Then, by Proposition \ref{prop:kl/phi-or-not-phi}$(ii)$, $(\Gamma_j, \Delta_j \cup \{\varphi_{j+1}\})$ has to be consistent and $\varphi_{j+1}$ is added by definition to $\Delta_j$. This implies $\varphi_{j+1} \in \Delta^*$, i.e.~$\neg\rho \in \Delta^*$.
   \item The proof for labeled formulas is the same as in the previous case and proceeds by contraposition by using Proposition \ref{prop:kl/phi-or-not-phi}$(i)$.   
   \end{enumerate}
\end{proof}

\begin{lemma}
	\label{lem:kl/properties-max-cons-pc}
	Let $(\Gamma, \Delta)$ be a maximally consistent proof context. Then:
	\begin{enumerate}[(i)]
	 \item $\; \Gamma, \Delta \vdash \varphi \,$ iff $\, \varphi \in (\Gamma, \Delta)$;
	 \item $\rho_1 \rimplies \rho_2 \, \in \, \Delta \;$ iff $\rho_1 \in \Delta$ implies $\rho_2 \in \Delta$;
   \item $\forall x.\, \rho \, \in \, \Delta \;$ iff $\rho[y/x] \in \Delta$ for all $y$;	 
	 \item $u: A\limplies B \, \in \, \Gamma \;$ iff $u:A \in \Gamma$ implies $u:B \in \Gamma$;
	 \item $u_1:\g A \, \in \, \Gamma \;$ iff $u_1 < u_2 \in \Delta$ implies $u_2:A \in \Gamma$ for all $u_2$;
	 \item $u_1:\h A \, \in \, \Gamma \;$ iff $u_2 < u_1 \in \Delta$ implies $u_2:A \in \Gamma$ for all $u_2$.
	\end{enumerate}
\end{lemma}
\begin{proof}
We treat only some cases, the others are similar and follow by maximality and consistency of $(\Gamma, \Delta)$.
  \begin{enumerate}	 
	 \item [$(i)$] The proof is analogous for rwffs and lwffs, we see the first case.
	 
	 $(\Leftarrow)$ If $\varphi \in (\Gamma, \Delta)$, then trivially $\Gamma, \Delta \vdash \varphi$.
	 
	 $(\Rightarrow)$ Consider an rwff $\varphi$ such that $\varphi \notin (\Gamma, \Delta)$. Then, by Definition \ref{def:kl/max-consistent-pc}, $\neg\varphi \in (\Gamma, \Delta)$. It follows trivially that $\Gamma, \Delta \vdash \neg\varphi$ holds. By hypothesis, $\Gamma, \Delta \vdash \varphi$ and thus by using $\neg E$ we get $\Gamma, \Delta \vdash \emptyset$, that contradicts the consistency of $(\Gamma, \Delta)$.
	 \item [$(v)$] $(\Leftarrow)$ Suppose $u_1:\g A \notin \Gamma$ and $u_2:A \in \Gamma$ for every $u_2$ such that $u_1<u_2 \in \Delta$. Then, by maximality of $(\Gamma, \Delta)$, $u_1:\non\g A \in \Gamma$. Now suppose there exists a $u_3$ such that $u_1 < u_3 \in \Delta$  and $u_3:\non A \in \Gamma$. Then, by hypothesis, we know $u_3:A \in \Gamma$ and this leads to a contradiction. Otherwise, if such a $u_3$ does not exist, we can conclude $u_1:\g A \in \Gamma$ that leads to a contradiction as well.
	 
	 $(\Rightarrow)$ We show it by contraposition. Suppose $u_1:\g A \in \Gamma$, $u_1<u_2 \in \Delta$ and $u_2:A \notin \Gamma$. By maximality of $(\Gamma, \Delta)$, we have $u_2:\non A \in \Gamma$. Then the following is an $\nkl$ proof that shows $(\Gamma, \Delta)$ is inconsistent.
	\end{enumerate}
	 \begin{prooftree}
    \axc{\scriptstyle u_1:\gp A}
		\axc{\scriptstyle u_1<u_2}
			\RightLabel{$\scriptstyle \gp E$}
		\bic{\scriptstyle u_2:A}
		\axc{\scriptstyle u_2:\non A}
			\RightLabel{$\scriptstyle \non E$}
		\bic{\scriptstyle u:\bottom}
   \end{prooftree}	 
\end{proof}

Our construction of maximally consistent proof contexts (Lemma \ref{lem:kl/max-consistent}) does not exclude the presence of two labels $x$ and $y$ that are related by the relation $x=y$. Now we want to derive a model from such a construction. Since we know from Definition \ref{def:kl/truth} that $\models^{\am,\lambda} x=y$ holds only if $\lambda(x)=\lambda(y)$,
we need to state an equivalence relation between labels on which the function $\lambda$ can be defined.
\begin{definition}
  \label{def:kl/equiv-c}
  Let $C = (\Gamma, \Delta)$ be a maximally consistent proof context and $L^C$ the set of labels occurring in it, we define the binary relation $\equiv^C$ on $L^C$ as follows: for every $u_1, u_2 \in L^C$, 
  \begin{center}
    $u_1 \equiv^C u_2$ \quad iff \quad $u_1 = u_2 \in \Delta$.
  \end{center}    
\end{definition}

\begin{proposition}
  \label{prop:kl/equiv-c}
  Given a maximally consistent proof context $C$, the relation $\equiv^C$ is an equivalence relation.
\end{proposition}
\begin{proof}
  It follows trivially by the maximality of $C$ and by the rules $\reflequal$, $\mon$, $\irreflless$ and $\mathit{conn}$.
\end{proof}

\begin{notation}
It follows from Proposition \ref{prop:kl/equiv-c} that every maximally consistent proof context $C$ determines a partition of the set $L^C$ of labels occurring in it. In the following, we will also use the notation $[u]^C$ to indicate the equivalence class containing the label $u$, i.e.
$$
  [u]^C = \{ u' \, \mid \, u \equiv^C u' \}.
$$
\end{notation}

\begin{definition}
	\label{def:kl/canonical-model}
	Let $C = (\Gamma, \Delta)$ be a maximally consistent proof context and $L^C$ be the set of labels occurring in it. We define the \emph{canonical model} $\m^C = (\wld^C, \prec^C, \val^C)$ as follows:
	\begin{itemize}
	 \item $\wld^C = \{ [u]^C \mid u \in L^C \}$; 
	 \item $([u_i]^C, [u_j]^C) \in \, \prec^C \;$ iff $\; u_i < u_j \in \Delta$;
	 \item $\val^C([u]^C,p) = 1 \;$ iff $\; u:p \in \Gamma$.
	\end{itemize}
	We define the \emph{canonical interpretation} $\lambda^C: L^C \rightarrow \wld^C$ as follows:
	 \begin{center}
	   $\lambda^C (u) = [u]^C$ for every $u \in L^C$.
	 \end{center}
\end{definition}

\begin{remark}
Note that in the previous definition $\prec^C$ and $\val^C$ are well defined, since it is easy to verify that for every $u_1, u_2 \in L^C$ it holds:
\begin{itemize}
\item $u_1 \equiv^C u_2$ implies for every $u_3 \in L^C$, $u_1<u_3 \in \Delta$ iff $u_2<u_3 \in \Delta$;
\item $u_1 \equiv^C u_2$ implies for every $u_3 \in L^C$, $u_3<u_1 \in \Delta$ iff $u_3<u_2 \in \Delta$;
\item $u_1 \equiv^C u_2$ implies for every $p \in \prop$, $u_1:p \in \Gamma$ iff $u_2:p \in \Gamma$.
\end{itemize}
\end{remark}

\begin{proposition}
	\label{prop:kl/canonical-model}
	Given a maximally consistent proof context $C = (\Gamma, \Delta)$, the canonical model $\m^C$ is a Kripke model for $\kl$.
\end{proposition}
\begin{proof}
	It suffices to show that $\m^C$ is irreflexive, transitive and connected.
	
	Suppose there exist three worlds $\mathcal{W}_1$, $\mathcal{W}_2$, and $\mathcal{W}_3$ in $\wld^C$ such that $(\mathcal{W}_1,\mathcal{W}_2) \in \prec^C$ and $(\mathcal{W}_2,\mathcal{W}_3) \in \prec^C$, but $(\mathcal{W}_1,\mathcal{W}_3) \notin \prec^C$. By definition \ref{def:kl/canonical-model}, this implies there exist at least three labels $x$, $y$ and $z$ such that $\lambda(x)= \mathcal{W}_1$, $\lambda(y) = \mathcal{W}_2$, $\lambda(z) = \mathcal{W}_3$, $x<y \in \Delta$ and $y<z \in \Delta$, but $x<z \notin \Delta$, i.e.~by the maximality of $C$, $\neg(x<z) \in \Delta$. But this leads to the inconsistency of $(\Gamma, \Delta)$, as shown by the following derivation.
	\begin{prooftree}
	\def\defaultHypSeparation{\hskip .0000001in}
	  \axc{}
	   \RightLabel{$\scriptstyle \transless$}
	 	\uic{\scriptstyle \forall x.y.z.\, (x<y \rand y<z) \rimplies x<z}
	 	 \RightLabel{$\scriptstyle \forall E$}
	 	 \doubleLine
	 	\uic{\scriptstyle (x<y \rand y<z) \rimplies x<z}
			\RightLabel{$\scriptstyle \transless$}
	 	\axc{\scriptstyle x<y}
	 	\axc{\scriptstyle y<z}
			\RightLabel{$\scriptstyle \rand I$}
    \bic{\scriptstyle x<y \rand y<z}
      \RightLabel{$\scriptstyle \rimpliesE$}
    \bic{\scriptstyle x<z}
		\axc{\scriptstyle \neg(x<z)}
			\RightLabel{$\scriptstyle \neg E$}
		\bic{\scriptstyle \emptyset}
 	\end{prooftree}

Connectedness of $\m^C$ can be proved in a similar way by using the rule $\mathit{conn}$. Suppose there exist two distinct worlds $\mathcal{W}_1$ and $\mathcal{W}_2$ in $\wld^C$ such that $(\mathcal{W}_1,\mathcal{W}_2) \notin \prec^C$ and $(\mathcal{W}_2,\mathcal{W}_1) \notin \prec^C$. By definition \ref{def:kl/canonical-model}, this implies there exist at least two labels $x$ and $y$ such that $\lambda(x)= \mathcal{W}_1$, $\lambda(y) = \mathcal{W}_2$, $x=y \notin \Delta$, $x<y \notin \Delta$ and $y<x \notin \Delta$, i.e. by the maximality of $C$, $\neg(x=y) \in \Delta$, $\neg(x<y) \in \Delta$ and $\neg(y<x) \in \Delta$. But this leads to the inconsistency of $(\Gamma, \Delta)$, as shown by the derivation in Figure~\ref{fig:conn-canonical}.

  \begin{figure*}[t]	
	 \begin{prooftree}

	  \axc{}
	   \RightLabel{$\scriptstyle \mathit{conn}$}
	 	\uic{\scriptstyle \forall x.y.\, x<y \ror x=y \ror y<x}
	   \RightLabel{$\scriptstyle \forall E$}
	   \doubleLine
	 	\uic{\scriptstyle x<y \ror x=y \ror y<x}
	 	
	 	\axc{\scriptstyle [x<y \ror x=y]^1}
	 	\axc{\scriptstyle [x<y]^2}
	 	\axc{\scriptstyle \neg(x<y)}
			\RightLabel{$\scriptstyle \neg E$}
	 	\bic{\scriptstyle \emptyset}
	 	\axc{\scriptstyle [x=y]^2}
	 	\axc{\scriptstyle \neg(x=y)}
			\RightLabel{$\scriptstyle \neg E$}
	 	\bic{\scriptstyle \emptyset}
			\RightLabel{$\scriptstyle \ror E^2$}
    \tic{\scriptstyle \emptyset}	 	
	 	
	 	\axc{\scriptstyle [y<x]^1}
	 	\axc{\scriptstyle \neg(y<x)}
			\RightLabel{$\scriptstyle \neg E$}
	 	\bic{\scriptstyle \emptyset}
	 	
			\RightLabel{$\scriptstyle \ror E^1$}
    \tic{\scriptstyle \emptyset}
	 		
	 \end{prooftree}
  \caption{Proof for connectedness of canonical models}\label{fig:conn-canonical}
	\end{figure*}

 	Irreflexivity of $\m^C$ can be shown in a similar way.  
\end{proof}

%
\begin{lemma}
	\label{lem:kl/completeness}
	Let $C=(\Gamma, \Delta)$ be a maximally consistent proof context, $\m^C$ the canonical model and $\lambda^C$ the canonical interpretation built on $C$ as in Definition~\ref{def:kl/canonical-model}. Then:
	\begin{enumerate}[(i)]
	 \item $\rho \in \Delta \;$ iff $\; \Gamma, \Delta \models^{\am^C, \lambda^C} \rho$;
	 \item $u:A \in \Gamma \;$ iff $\; \Gamma, \Delta \models^{\am^C, \lambda^C} u:A$.
	\end{enumerate}
\end{lemma}
\begin{proof}
	\begin{enumerate}[(i)]
\item $(\Rightarrow)$ By hypothesis, $\rho \in \Delta$. Then, if we assume $\models^{\am^C, \lambda^C} (\Gamma,\Delta)$, it immediately follows $\models^{\am^C, \lambda^C} \rho$.

 	 $(\Leftarrow)$ By hypothesis, $\Gamma, \Delta \models^{\am^C, \lambda^C} \rho$. Let us suppose $\rho \notin \Delta$. By maximality of $(\Gamma, \Delta)$, it follows $\neg \rho \in \Delta$. Then we have also $\Gamma, \Delta \models^{\am^C, \lambda^C} \neg \rho$ (see direction ($\Rightarrow$)). But, since we have by hypothesis $\Gamma, \Delta \models^{\am^C, \lambda^C} \rho$, this yields the absurd $\Gamma, \Delta \models^{\am^C, \lambda^C} \emptyset$.
\item The proof for labeled formulas is analogous.	
	\end{enumerate}
\end{proof}

\begin{theorem}
	\label{th:kl/nkl-completeness}
	$\nkl = \nl + \nr + \ngen$ is complete, i.e. it holds:
	\begin{enumerate}[(i)]
	 \item if $\Gamma, \Delta \nvdash w:A \;$, then there exist a $\kl$ model $\m^C$ and an interpretation $\lambda^C$ such that $\Gamma, \Delta \nmodels^{\am^C,\lambda^C} w:A$;
	 \item if $\Gamma, \Delta \nvdash \rho \;$, then there exist a $\kl$ model $\m^C$ and an interpretation $\lambda^C$ such that $\Gamma, \Delta \nmodels^{\am^C,\lambda^C} \rho$.
	\end{enumerate}
\end{theorem}
\begin{proof}
	\begin{enumerate}[(i)]
	 \item If $\Gamma, \Delta \nvdash w:A$, then $(\Gamma \cup \{ w:\non A \}, \Delta)$ is consistent; otherwise there exists a $w_i$ such that $\Gamma \cup \{ w: \non A\}, \Delta \vdash w_i:\bottom$, and then $\Gamma, \Delta \vdash w:A$. Therefore, by Lemma \ref{lem:kl/max-consistent}, $(\Gamma \cup \{ w:\non A \}, \Delta)$ is included in a maximally consistent proof context $C = ((\Gamma \cup \{ w:\non A \})^*, \Delta^*)$. Let $\m^C$ be the canonical model for $C$. It suffices to find an interpretation according to which $\m^C$ is not a model for $w:A$. By Lemma \ref{lem:kl/completeness}, $(\Gamma \cup \{ w:\non A \})^*, \Delta^* \models^{\am^C,\lambda^C} w:\non A$, where $\m^C$ is a $\kl$ model by Proposition \ref{prop:kl/canonical-model}. It follows $\Gamma \cup \{ w:\non A \})^*, \Delta^* \nmodels^{\am^C,\lambda^C} w: A$, and thus $\Gamma, \Delta \nmodels^{\am^C,\lambda^C} w:A$.
	 \item We can repeat the same proof for relational formulas. If $\Gamma, \Delta \nvdash \rho$, then $(\Gamma, \Delta \cup \{ \neg\rho \})$ is consistent. Then we can build a maximally consistent proof context $\Gamma^*, (\Delta \cup \{ \neg\rho \})^*$ such that $\Gamma^*, (\Delta \cup \{ \neg\rho \})^* \nmodels^{\am^C, \lambda^C} \rho$, and thus $\Gamma, \Delta \nmodels^{\am^C, \lambda^C} \rho$.
	\end{enumerate}
\end{proof}

\subsubsection{Completeness by axioms}\label{sec:completeness-axioms}
  It is possible to give an indirect proof of completeness by showing that all the axioms listed in \secref{subsub:axiomatization} for the logic $\kl$ are derivable in $\nkl$.
  In the following derivations, for simplicity, we will sometimes use derived operators and derived rules, and exploit trivial equivalences between formulas implicitly.

We begin by giving derivations for the axioms $\mathit(G1)$ and $\mathit{(G2)}$:
\begin{displaymath}
    \footnotesize
\renewcommand{\arraystretch}{6}
    \begin{array}{c}
          
    \axc{\scriptstyle [t:\gp (A \limplies B)]^1}
    \axc{\scriptstyle [t<s]^3}
	   \RightLabel{$\scriptstyle \gp E$}
	 	\bic{\scriptstyle s:A \limplies B}    
    
    \axc{\scriptstyle [t:\gp A]^2}
    \axc{\scriptstyle [t<s]^3}
	   \RightLabel{$\scriptstyle \gp E$}
	 	\bic{\scriptstyle s:A}
    
	   \RightLabel{$\scriptstyle \limplies E$}
	 	\bic{\scriptstyle s:B}
    
	   \RightLabel{$\scriptstyle \gp I^3$}
	 	\uic{\scriptstyle t:\gp B}
    
	   \RightLabel{$\scriptstyle \limplies I^2$}
	 	\uic{\scriptstyle t:\gp A \limplies \gp B}

	   \RightLabel{$\scriptstyle \limplies I^1$}
	 	\uic{\scriptstyle t:\gp (A\limplies B) \limplies (\gp A \limplies \gp B)}
    
    \DisplayProof 

\\

          
    \axc{\scriptstyle [t:\pp \gp A]^1}
    
    \axc{\scriptstyle [s:\gp A]^2}
    \axc{\scriptstyle [s<t]^2}
	   \RightLabel{$\scriptstyle \gp E$}
	 	\bic{\scriptstyle t:A}

	   \RightLabel{$\scriptstyle \pp E^2$}
	 	\bic{\scriptstyle t:A}
	 	
	   \RightLabel{$\scriptstyle \limplies I^1$}
	 	\uic{\scriptstyle t:\pp \gp A \limplies A}
    
    \DisplayProof  

  \end{array}
\end{displaymath}
The derivation for $\mathit{(G3)}$ is shown in Figure~\ref{fig:axiom-G3}, while the derivation for $\mathit{(G4)}$ is in Figure~\ref{fig:axiom-G4}.  We omit here the derivations for the symmetric axioms $\mathit{(H1)}$-$\mathit{(H4)}$.

\begin{figure*}\footnotesize
\begin{center}
    \footnotesize

    \axc{\scriptstyle [t:\gp A]^1}

	  \axc{}
	   \RightLabel{$\scriptstyle \transless$}
	 	\uic{\scriptstyle \forall x.y.z.\, (x<y \rand y<z) \rimplies x<z}
	   \RightLabel{$\scriptstyle \forall E$}
	 	\uic{\scriptstyle \forall y.z.\, (t<y \rand y<z) \rimplies t<z}
	   \RightLabel{$\scriptstyle \forall E$}
	 	\uic{\scriptstyle \forall z.\, (t<s \rand s<z) \rimplies t<z}
	   \RightLabel{$\scriptstyle \forall E$}	 	
	 	\uic{\scriptstyle (t<s \rand s<r) \rimplies t<r}

    \axc{\scriptstyle [t<s]^2}
    \axc{\scriptstyle [s<r]^3}
	   \RightLabel{$\scriptstyle \rand I$}
	 	\bic{\scriptstyle t<s \, \rand \, s<r}

	   \RightLabel{$\scriptstyle \rimplies E$}
	 	\bic{\scriptstyle t<r}

	   \RightLabel{$\scriptstyle \gp E$}
	 	\bic{\scriptstyle r:A}

	   \RightLabel{$\scriptstyle \gp I^3$}
	 	\uic{\scriptstyle s:\gp A}
	 	
	   \RightLabel{$\scriptstyle \gp I^2$}
	 	\uic{\scriptstyle t:\gp \gp A}
	 	
	   \RightLabel{$\scriptstyle \limplies I^1$}
	 	\uic{\scriptstyle t:\gp A \limplies \gp \gp A}

    \DisplayProof
    
    \end{center}
    
  \caption{Derivation of the axiom $\mathit{(G3)}$}
  \label{fig:axiom-G3}
\end{figure*}
\begin{figure*}\footnotesize
\begin{center}
	  \def\defaultHypSeparation{\hskip .2in}
    \axc{\scriptstyle [t:\fp \non A \wedge \fp \non B]^2}
	   \RightLabel{$\scriptstyle \wedge E$}
	 	\uic{\scriptstyle t:\fp \non A}

    \axc{\scriptstyle [t:\fp \non A \wedge \fp \non B]^2}
	   \RightLabel{$\scriptstyle \wedge E$}
	 	\uic{\scriptstyle t:\fp \non B}
	 		  	
	  \axc{}
	   \RightLabel{$\scriptstyle \mathit{conn}$}
	 	\uic{\scriptstyle \forall x.y.\, x<y \ror x=y \ror y<x}
	   \RightLabel{$\scriptstyle \forall E$}
	 	\uic{\scriptstyle \forall y.\, s<y \ror s=y \ror y<s}
	   \RightLabel{$\scriptstyle \forall E$}
	 	\uic{\scriptstyle s<r \ror s=r \ror r<s}

	 	\axc{\pi_1}
	 	 \noLine
	 	\uic{\scriptstyle \emptyset}

	 	\axc{\scriptstyle [s=r \ror r<s]^5}
	 	
	 	\axc{\pi_2}
	 	 \noLine
	 	\uic{\scriptstyle \emptyset}
	 	
	 	\axc{\pi_3}
	 	 \noLine
	 	\uic{\scriptstyle \emptyset}	 	
	 	
	   \RightLabel{$\scriptstyle \ror E^8$}
	 	\tic{\scriptstyle \emptyset}

	   \RightLabel{$\scriptstyle \ror E^5$}
	 	\tic{\scriptstyle \emptyset}
	 	
	   \RightLabel{$\scriptstyle \mathit{uf}2$}
	 	\uic{\scriptstyle t:\bottom}
	 	
	   \RightLabel{$\scriptstyle \fp E^4$}
	 	\bic{\scriptstyle t:\bottom}
	 		 	
	   \RightLabel{$\scriptstyle \fp E^3$}
	 	\bic{\scriptstyle t:\bottom}

	   \RightLabel{$\scriptstyle {\bottomE}^2$}
	 	\uic{\scriptstyle t:\gp A \vee \gp B}
	 	
	 	 \RightLabel{$\scriptstyle \limplies I^1$}
	 	\uic{\scriptstyle t:(\gp (A \vee B) \wedge \gp (A \vee \gp B) \wedge \gp (\gp A \vee B)) \limplies (\gp A \vee \gp B)}
\DisplayProof
\end{center}
\bigskip
    where $\pi_1$ is:
\smallskip

	  \def\defaultHypSeparation{\hskip .05in}
    \axc{\scriptstyle [r:\non B]^4}

	 	\axc{\scriptstyle [t:(\gp (A \vee B) \wedge \gp (A \vee \gp B) \wedge \gp (\gp A \vee B))]^1}
	   \RightLabel{$\scriptstyle \wedge E$}
	 	\uic{\scriptstyle t:\gp (A \vee \gp B)}

    \axc{\scriptstyle [t<s]^3}
	   \RightLabel{$\scriptstyle \gp E$}
	 	\bic{\scriptstyle (s:A \vee \gp B)}

	 	\axc{\scriptstyle [s:\non A]^3}
	 	\axc{\scriptstyle [s: A]^7}
	   \RightLabel{$\scriptstyle \non E$}
	 	\bic{\scriptstyle s:\bottom}

	 	\axc{\scriptstyle [s:\non \gp B]^6}
	 	\axc{\scriptstyle [s:\gp B]^7}
	   \RightLabel{$\scriptstyle \non E$}
	 	\bic{\scriptstyle s:\bottom}

	   \RightLabel{$\scriptstyle \vee E^7$}
    \tic{\scriptstyle s:\bottom}

	   \RightLabel{$\scriptstyle {\bottomE}^6$}
	 	\uic{\scriptstyle s:\gp B}

    \axc{\scriptstyle [s<r]^5}
     \RightLabel{$\scriptstyle \g E$}
    \bic{\scriptstyle r:B}

     \RightLabel{$\scriptstyle \non E$}
    \bic{\scriptstyle r:\bottom}

     \RightLabel{$\scriptstyle \mathit{uf}1$}
    \uic{\scriptstyle \emptyset}    
	 	\DisplayProof      

\bigskip
    $\pi_2$ is:
\smallskip

	  \def\defaultHypSeparation{\hskip .05in}

    \axc{\scriptstyle [r:\non B]^4}

	 	\axc{\scriptstyle [t:(\gp (A \vee B) \wedge \gp (A \vee \gp B) \wedge \gp (\gp A \vee B))]^1}
	   \RightLabel{$\scriptstyle \wedge E$}
	 	\uic{\scriptstyle t:\gp (A \vee B)}

	 	\axc{\scriptstyle [t<s]^3}
	   \RightLabel{$\scriptstyle \gp E$}
	 	\bic{\scriptstyle s: A \vee B}
	 	
	 	\axc{\scriptstyle [s:\non A]^3}
	 	\axc{\scriptstyle [s:A]^{12}}
	   \RightLabel{$\scriptstyle \non E$}
	 	\bic{\scriptstyle s: \bottom}

	 	\axc{\scriptstyle [s:\non B]^{11}}
	 	\axc{\scriptstyle [s:B]^{12}}
	   \RightLabel{$\scriptstyle \non E$}
	 	\bic{\scriptstyle s: \bottom}

	   \RightLabel{$\scriptstyle \vee E^{12}$}
	 	\tic{\scriptstyle s:\bottom}
	 	
	   \RightLabel{$\scriptstyle {\bottomE}^{11}$}
	 	\uic{\scriptstyle s:B}
	 	
	 	\axc{\scriptstyle [s=r]^8}
	   \RightLabel{$\scriptstyle \mon$}
	 	\bic{\scriptstyle r:B}

	   \RightLabel{$\scriptstyle \non E$}
	 	\bic{\scriptstyle r: \bottom}

	   \RightLabel{$\scriptstyle \mathit{uf}1$}
	 	\uic{\scriptstyle \emptyset}
    \DisplayProof

\bigskip
    and $\pi_3$ is:
\smallskip

	  \def\defaultHypSeparation{\hskip .05in}

    \axc{\scriptstyle [s:\non A]^3}	 	
	 	
	 	\axc{\scriptstyle [t:(\gp (A \vee B) \wedge \gp (A \vee \gp B) \wedge \gp (\gp A \vee B))]^1}
	   \RightLabel{$\scriptstyle \wedge E$}
	 	\uic{\scriptstyle t:\gp (\gp A \vee B)}

	 	\axc{\scriptstyle [t<r]^4}
	   \RightLabel{$\scriptstyle \gp E$}
	 	\bic{\scriptstyle r: \gp A \vee B}
	 	
	 	\axc{\scriptstyle [r:\non \gp A]^9}
	 	\axc{\scriptstyle [r:\gp A]^{10}}
	   \RightLabel{$\scriptstyle \non E$}
	 	\bic{\scriptstyle r: \bottom}

	 	\axc{\scriptstyle [r:\non B]^{4}}
	 	\axc{\scriptstyle [r:B]^{10}}
	   \RightLabel{$\scriptstyle \non E$}
	 	\bic{\scriptstyle r: \bottom}

	   \RightLabel{$\scriptstyle \vee E^{10}$}
	 	\tic{\scriptstyle r:\bottom}
	 	
	   \RightLabel{$\scriptstyle {\bottomE}^{9}$}
	 	\uic{\scriptstyle r:\gp A}
	 	
	 	\axc{\scriptstyle [r<s]^8}
	   \RightLabel{$\scriptstyle \gp E$}
	 	\bic{\scriptstyle s:A}

	   \RightLabel{$\scriptstyle \non E$}
	 	\bic{\scriptstyle s: \bottom}

	   \RightLabel{$\scriptstyle \mathit{uf}1$}
	 	\uic{\scriptstyle \emptyset}
    \DisplayProof

\caption{Derivation of the axiom $\mathit{(G4)}$}
\label{fig:axiom-G4}
\end{figure*}

  Completeness of the extended systems considered in \secref{sec:family} can be also proved by deriving the corresponding axioms. In~\secref{sec:family}, we have already proved the axioms for \emph{having a first point} and \emph{right-seriality}. We show the derivations for \emph{right-density} and for \emph{right-discreteness} in Figure~\ref{fig:axiom-rdensity} and Figure~\ref{fig:axiom-rdiscreteness}, respectively. Derivations of the other axioms (\emph{final point}, \emph{left-seriality}, \emph{left-density}, \emph{left-discreteness}) are symmetric and we thus omit them.

				  \begin{figure*} 
				  \axc{\scriptstyle [t: \fp A]^1}
				  
				  \axc{}
  					\RightLabel{$\scriptstyle \mathit{dens}$}
					\uic{\scriptstyle \forall x.y.\, x<y \rimplies \exists z.\, (x<z \rand z<y)}
  					\RightLabel{$\scriptstyle \forall E$}
					\uic{\scriptstyle \forall y.\, (t<y \rimplies \exists z.\, (t<z \rand z<y))}
  					\RightLabel{$\scriptstyle \forall E$}
					\uic{\scriptstyle t<s \rimplies \exists z.\, (t<z \rand z<s)}
					\axc{\scriptstyle [t<s]^2}
  					\RightLabel{$\scriptstyle \rimplies E$}
					\bic{\scriptstyle \exists z.\, (t<z \rand z<s)}
					
					\axc{\scriptstyle [t:\non \fp \fp A]^3}
					\axc{\scriptstyle [s: A]^2}
					\axc{\scriptstyle [t<r \rand r<s]^4}
  					\RightLabel{$\scriptstyle \rand E$}
					\uic{\scriptstyle r<s}
  					\RightLabel{$\scriptstyle \fp I$}					
					\bic{\scriptstyle r: \fp A}

					\axc{\scriptstyle [t<r \rand r<s]^4}
  					\RightLabel{$\scriptstyle \rand E$}
					\uic{\scriptstyle t<r}
  					\RightLabel{$\scriptstyle \fp I$}
					\bic{\scriptstyle t: \fp \fp A}
  					\RightLabel{$\scriptstyle \non E$}
					\bic{\scriptstyle t: \bottom}

  					\RightLabel{$\scriptstyle \mathit{uf}1$}
					\uic{\scriptstyle \emptyset}

					  \RightLabel{$\scriptstyle \exists E^4$}
					\bic{\scriptstyle \emptyset}

					  \RightLabel{$\scriptstyle \mathit{uf}2$}
					\uic{\scriptstyle t:\bottom}
					  \RightLabel{$\scriptstyle {\bottomE}^3$}
					\uic{\scriptstyle t:\fp \fp A}
					  \RightLabel{$\scriptstyle \fp E^2$}
					\bic{\scriptstyle t:\fp \fp A}
					  \RightLabel{$\scriptstyle \limplies I^1$}
					\uic{\scriptstyle t:\fp A \limplies \fp \fp A}

					\DisplayProof
			    \caption{Derivation of the modal axiom for right-density{\label{fig:axiom-rdensity}}}

				 \end{figure*}
				 
\begin{figure*}
    \footnotesize
	  \def\defaultHypSeparation{\hskip .05in}

    \axc{\scriptstyle [t:\fp \true \wedge A \wedge \hp A]^1}
	   \RightLabel{$\scriptstyle \wedge E$}
	 	\uic{\scriptstyle t:\fp \true}
	 	
	  \axc{}
	   \RightLabel{$\scriptstyle \mathit{rdiscr}$}
	 	\uic{\scriptstyle \forall x.y.\, x<y \rimplies (\exists z. \, x<z \rand (\neg \exists u.\, x<u \rand u<z))}
	   \RightLabel{$\scriptstyle \forall E$}
	 	\uic{\scriptstyle \forall y.\, t<y \rimplies (\exists z. \, t<z \rand (\forall u.\, \neg(t<u) \ror \neg (u<z)))}
	   \RightLabel{$\scriptstyle \forall E$}
	 	\uic{\scriptstyle t<q \rimplies (\exists z. \, t<z \rand (\forall u.\, \neg(t<u) \ror \neg (u<z)))}
	   
	  \axc{\scriptstyle [t<u]^3}
	   \RightLabel{$\scriptstyle \rimplies E$}
	 	\bic{\scriptstyle \exists z. \, t<z \rand (\forall u.\, \neg(t<u) \ror \neg (u<z))}	
	 	   
	  \axc{\scriptstyle [t: \non \fp \hp A]^2}	 	   	 	   
	   
	  \axc{\pi}
	 	 \noLine
	 	\uic{\scriptstyle \emptyset}	   

	   \RightLabel{$\scriptstyle \mathit{uf}2$}
	 	\uic{\scriptstyle r:\bottom}

	   \RightLabel{$\scriptstyle {\bottomE}^6$}
	 	\uic{\scriptstyle r:A}

	   \RightLabel{$\scriptstyle \hp I^5$}
	 	\uic{\scriptstyle s:\hp A}

	  \axc{\scriptstyle [t<s \rand (\forall u. \, \neg(t<u) \ror \neg (u<s))]^4}
	   \RightLabel{$\scriptstyle \rand E$}
	 	\uic{\scriptstyle t<s}

	   \RightLabel{$\scriptstyle \fp I$}
	 	\bic{\scriptstyle t:\fp \hp A}

	   \RightLabel{$\scriptstyle \non E$}
	 	\bic{\scriptstyle t:\bottom}

	   \RightLabel{$\scriptstyle \mathit{uf}1$}
	 	\uic{\scriptstyle \emptyset}

	   \RightLabel{$\scriptstyle \exists E^4$}
	 	\bic{\scriptstyle \emptyset}

	   \RightLabel{$\scriptstyle \fp E^3$}
	 	\bic{\scriptstyle \emptyset}

	   \RightLabel{$\scriptstyle \mathit{uf}2$}
	 	\uic{\scriptstyle t:\bottom}

	   \RightLabel{$\scriptstyle {\bottomE}^2$}
	 	\uic{\scriptstyle t:\fp \hp A}

	   \RightLabel{$\scriptstyle \limplies I^1$}
	 	\uic{\scriptstyle t:(\fp \true \wedge A \wedge \hp A) \limplies \fp \hp A}
	 	
    \DisplayProof
    
\bigskip
where $\pi$ is:
\smallskip
	  \def\defaultHypSeparation{\hskip .05in}
	  \axc{}
	   \RightLabel{$\scriptstyle \mathit{conn}$}
	 	\uic{\scriptstyle \forall x.y.\, x<y \ror x=y \ror y<x}
	   \RightLabel{$\scriptstyle \forall E$}
	 	\uic{\scriptstyle \forall y.\, r<y \ror r=y \ror y<r}
	   \RightLabel{$\scriptstyle \forall E$}
	 	\uic{\scriptstyle r<t \ror r=t \ror t<r}
	 	
	  \axc{\scriptstyle [r:\non A]^6}

    \axc{\scriptstyle [t:\fp \true \wedge A \wedge \hp A]^1}
	   \RightLabel{$\scriptstyle \wedge E$}
	 	\uic{\scriptstyle t:\hp A}
	 	
	  \axc{\scriptstyle [r<t]^7}
	   \RightLabel{$\scriptstyle \hp E$}
	 	\bic{\scriptstyle r: A}
	  
	   \RightLabel{$\scriptstyle \non E$}
	 	\bic{\scriptstyle r: \bottom}
	   
	   \RightLabel{$\scriptstyle \mathit{uf}1$}
	 	\uic{\scriptstyle \emptyset}	   
    
	 	\axc{\pi_1}
	 	 \noLine
	 	\uic{\scriptstyle \emptyset}

	   \RightLabel{$\scriptstyle \ror E^7$}	   
	 	\tic{\scriptstyle \emptyset}    
    
    \DisplayProof
    
\bigskip
and $\pi_1$ is:
\smallskip
	  \def\defaultHypSeparation{\hskip .05in}
	  \axc{\scriptstyle [r=t \ror t<r]^7}
	  
	  \axc{\scriptstyle [r:\non A]^6}
	  
	  \axc{\scriptstyle [t:\fp \true \wedge A \wedge \hp A]^1}
	   \RightLabel{$\scriptstyle \wedge E$}
	 	\uic{\scriptstyle t: A}
	  
	  \axc{\scriptstyle [r=t]^8}
	   \RightLabel{$\scriptstyle \mon$}
	 	\bic{\scriptstyle r: A}
	  
	   \RightLabel{$\scriptstyle \non E$}
	 	\bic{\scriptstyle r: \bottom}
	 	
	   \RightLabel{$\scriptstyle \mathit{uf}1$}
	 	\uic{\scriptstyle \emptyset}
	 	
	  \axc{\scriptstyle [t<s \rand (\forall u. \neg(t<u) \ror \neg(u<s))]^4}
	   \RightLabel{$\scriptstyle \rand E$}
	 	\uic{\scriptstyle \forall u. \neg(t<u) \ror \neg(u<s)}
	 	
	   \RightLabel{$\scriptstyle \forall E$}
	 	\uic{\scriptstyle \neg(t<r) \ror \neg(r<s)}
	 	
	  \axc{\scriptstyle [\neg(t<r)]^9}
	  \axc{\scriptstyle [t<r]^8}
	   \RightLabel{$\scriptstyle \neg E$}
	 	\bic{\scriptstyle \emptyset}
	 	
	 	\axc{\scriptstyle [\neg(r<s)]^9}
	  \axc{\scriptstyle [r<s]^5}
	   \RightLabel{$\scriptstyle \neg E$}
	 	\bic{\scriptstyle \emptyset}
	  
	   \RightLabel{$\scriptstyle \ror E^9$}
	 	\tic{\scriptstyle \emptyset}
	  
	   \RightLabel{$\scriptstyle \ror E^8$}
	 	\tic{\scriptstyle \emptyset}
    
    \DisplayProof    
    
\caption{Derivation of the modal axiom for right-discreteness}
\label{fig:axiom-rdiscreteness}
\end{figure*}

\subsection{Normalization}
  \label{sub:appendix-normalization}

\begin{proof}[Lemma~\ref{lem:kl/restrictions}]\\
$(i)$ We show that any application of $\bottomE$, $\emptysetE$, and $\mon$ with a non-atomic conclusion can be replaced with a derivation in which such rules are applied only to formulas of smaller grade by the set of transformations given below. By iterating these transformations, we get a derivation of $\varphi$ from $\Gamma, \Delta$ where the conclusions of applications of $\bottomE$, $\emptysetE$, and $\mon$ are atomic.

(1) First, we consider applications of $\bottomE$. There are three possible cases, depending on whether the conclusion is $x:B\limplies C$, $x:\g B$, or $x:\h B$. Note that in the following transformations we only show the part of the derivation where the reduction, denoted by $\rightsquigarrow$, actually takes place; the missing parts remain unchanged.
$$
	  \begin{array}{lcc}
	  \\
    \emph{(Case 1)}
    &&
    
    \\	  
	  \def\defaultHypSeparation{\hskip .00001in}	  
	  \axc{\scriptstyle [x:(B\limplies C)\limplies \bottom]}
	   \noLine
	 	\uic{\scriptstyle \pi}
	 	 \noLine
	 	\uic{\scriptstyle y:\bottom}
			\RightLabel{$\scriptstyle \bottomE$}
	 	\uic{\scriptstyle x:B\limplies C}
	 \DisplayProof
	 
	    &
	  \rightsquigarrow
	    &
	    
\def\defaultHypSeparation{\hskip .00001in}
	  \axc{\scriptstyle [x:C\limplies \bottom]^2}
	  \axc{\scriptstyle [x:B\limplies C]^1}
	  \axc{\scriptstyle [x:B]^3}
	   \RightLabel{$\scriptstyle \limplies E$}
	  \bic{\scriptstyle x:C}
	   \RightLabel{$\scriptstyle \limplies E$}
	  \bic{\scriptstyle x:\bottom}
	   \RightLabel{$\scriptstyle \limplies I^1$}
	  \uic{\scriptstyle x:(B\limplies C) \limplies \bottom}	  
	   \noLine
	  \uic{\scriptstyle \pi}
	   \noLine
	  \uic{\scriptstyle y:\bottom}
	   \RightLabel{$\scriptstyle {\bottomE}^2$}
	  \uic{\scriptstyle x:C}
	   \RightLabel{$\scriptstyle \limplies I^3$}
	  \uic{\scriptstyle x:B\limplies C}
	  \DisplayProof
	  
	  \end{array}
$$
$$	  
	  \begin{array}{lcc}
	  	  \\
	      \emph{(Case 2)}
    &&
    
    \\	  
\def\defaultHypSeparation{\hskip .00001in}	  
	  \axc{\scriptstyle [x:\gp B \limplies \bottom]}
	   \noLine
	 	\uic{\scriptstyle \pi}
	 	 \noLine
	 	\uic{\scriptstyle y:\bottom}
			\RightLabel{$\scriptstyle \bottomE$}
	 	\uic{\scriptstyle x:\gp B}
	 \DisplayProof	 
	    &
	  \rightsquigarrow
	    &
\def\defaultHypSeparation{\hskip .00001in}	    
	  \axc{\scriptstyle [y:B\limplies \bottom]^2}
	  \axc{\scriptstyle [x:\gp B]^1}
	  \axc{\scriptstyle [x<y]^3}
	   \RightLabel{$\scriptstyle \gp E$}
	  \bic{\scriptstyle y:B}
	   \RightLabel{$\scriptstyle \limplies E$}
	  \bic{\scriptstyle y:\bottom}
	   \RightLabel{$\scriptstyle \bottomE$}
	  \uic{\scriptstyle x:\bottom}	  
	   \RightLabel{$\scriptstyle \limplies I^1$}
	  \uic{\scriptstyle x:\gp B \limplies \bottom}	  
	   \noLine
	  \uic{\scriptstyle \pi}
	   \noLine
	  \uic{\scriptstyle y:\bottom}
	   \RightLabel{$\scriptstyle {\bottomE}^2$}
	  \uic{\scriptstyle y:B}
	   \RightLabel{$\scriptstyle  \gp I^3$}	  
	  \uic{\scriptstyle x:\gp B}
	   \DisplayProof
  
	  \end{array}
$$
Case 3 concerns formulas of the form $y:\h A$; it is analogous to the previous one and we omit the reduction for it.

(2) Applications of $\emptysetE$ can be reduced to applications on formulas of lower grade, following an approach analogous to that of $\bottomE$. It is easy to see that in this case, we can also restrict to applications of $\emptysetE$ in which the conclusion is not $\emptyset$. We have to consider two possibilities: formulas of the form $\rho_1 \rimplies \rho_2$ and formulas of the form $\forall x.\, \rho$. We consider only the second case, since the first one is analogous to the case of implication for labeled formulas:
$$
	  \begin{array}{ccc}

	  \axc{\scriptstyle [\forall x.\, \rho \rimplies \emptyset]}
	   \noLine
	 	\uic{\scriptstyle \pi}
	 	 \noLine
	 	\uic{\scriptstyle \emptyset}
			\RightLabel{$\scriptstyle \emptysetE$}
	 	\uic{\scriptstyle \forall x.\, \rho}
	 \DisplayProof
	 
	    &
	  \rightsquigarrow
	    &
	    
	  \axc{\scriptstyle  [\rho \rimplies \emptyset]^1}
	     \RightLabel{$\scriptstyle \forall I$}
	  \uic{\scriptstyle \forall x.\, \rho \rimplies \emptyset}
	   \noLine
	 	\uic{\scriptstyle \pi}
	 	 \noLine
	  \uic{\scriptstyle \emptyset}
	   \RightLabel{$\scriptstyle \emptysetE^1$}
	  \uic{\scriptstyle \rho}
	   \RightLabel{$\scriptstyle \forall I$}
	  \uic{\scriptstyle \forall x.\, \rho}
	   \DisplayProof

	  \end{array}
$$

(3) Finally, we consider applications of the rule $\mon$. We have five cases depending on the form of the formula that is the major premise of the $\mon$ application:
  \begin{enumerate}[(a)]
    \item $x:A\limplies B$
    \item $x:\g A$
    \item $x:\h A$
    \item $\rho_1 \rimplies \rho_2$
    \item $\forall x.\, \rho$
  \end{enumerate}
$$
	  \begin{array}{lcc}
	  \\
	  
    \emph{(Case a)}
    &&
    
    \\	  
\def\defaultHypSeparation{\hskip .00000001in}	  
	  \axc{\scriptstyle x:A\limplies B}
	  \axc{\scriptstyle x=y}
			\RightLabel{$\scriptstyle \mon$}
	 	\bic{\scriptstyle y:A\limplies B}
	 \DisplayProof
	 
	    &
	  \rightsquigarrow
	    &
\def\defaultHypSeparation{\hskip .00000001in}	    
	  \axc{\scriptstyle x:A\limplies B}
	  \axc{\scriptstyle [y:A]^1}
	  \axc{\scriptstyle x=y}
			\RightLabel{$\scriptstyle \mon$}
	 	\bic{\scriptstyle x:A}
			\RightLabel{$\scriptstyle \limplies E$}	 	
	 	\bic{\scriptstyle x:B}
	 	\axc{\scriptstyle x=y}
	 	  \RightLabel{$\scriptstyle \mon$}
    \bic{\scriptstyle y:B}
	 	  \RightLabel{$\scriptstyle \limplies I^1$}    
	 	\uic{\scriptstyle y:A\limplies B}
	 \DisplayProof	    
	  
	  \end{array}
$$
$$
    \begin{array}{lcc}	  
	      \emph{(Case b)}
    &&
    
    \\	  
\def\defaultHypSeparation{\hskip .00000001in}  
	  \axc{\scriptstyle x:\gp A}
	  \axc{\scriptstyle x=y}
			\RightLabel{$\scriptstyle \mon$}
	 	\bic{\scriptstyle y:\gp A}
	 \DisplayProof
	 
	    &
	  \rightsquigarrow
	    &
\def\defaultHypSeparation{\hskip .00000001in}	    
	  \axc{\scriptstyle x:\gp A}
	  \axc{\scriptstyle [y<z]^1}
	  \axc{\scriptstyle x=y}
			\RightLabel{$\scriptstyle \mon$}
	 	\bic{\scriptstyle x<z}
			\RightLabel{$\scriptstyle \gp E$}	 	
	 	\bic{\scriptstyle z:A}
	 	  \RightLabel{$\scriptstyle \gp I^1$}
    \uic{\scriptstyle y:\gp A}
	 \DisplayProof	    
  
	  \end{array}
$$  
$$
    \begin{array}{lcc}	  
	      \emph{(Case e)}
    &&
    
    \\	  
\def\defaultHypSeparation{\hskip .00000001in}  
	  \axc{\scriptstyle \forall x. \, \rho}
	  \axc{\scriptstyle y=z}
			\RightLabel{$\scriptstyle \mon$}
	 	\bic{\scriptstyle \forall x. \, \rho[z/y]}
	 \DisplayProof
	 
	    &
	  \rightsquigarrow
	    &
\def\defaultHypSeparation{\hskip .00000001in}	    
	  \axc{\scriptstyle \forall x. \, \rho}
	   \RightLabel{$\scriptstyle \forall E$}
	  \uic{\scriptstyle \rho}
	  \axc{\scriptstyle y=z}
			\RightLabel{$\scriptstyle \mon$}
	 	\bic{\scriptstyle \rho[z/y]}
			\RightLabel{$\scriptstyle \forall I$}	 	
	 	\uic{\scriptstyle \forall x. \, \rho[z/y]}
	 \DisplayProof	    
  
	  \end{array}
$$  
The case $(c)$ is analogous to $(b)$, while the transformation for the case $(d)$ is as in $(a)$ where $\rimplies$ plays the role of $\limplies$.
  
$(ii)$ We show that every application of $\mon$ on a lwff of the form $x:\bottom$ can be replaced by an application of $\bottomE$ that does not discharge any assumption:
$$
    \begin{array}{ccc}	  
    \alwaysNoLine
	  \axc{\scriptstyle \pi}
	  \uic{\scriptstyle x:\bottom}

	  \axc{\scriptstyle \pi'}
	  \uic{\scriptstyle x=y}
	   \alwaysSingleLine
			\RightLabel{$\scriptstyle \mon$}
	 	\bic{\scriptstyle y:\bottom}
	 \DisplayProof
	    &
	  \rightsquigarrow
	    &
    \alwaysNoLine
	  \axc{\scriptstyle \pi}
	  \uic{\scriptstyle x:\bottom}
	   \singleLine
			\RightLabel{$\scriptstyle \bottomE$}
	  \uic{\scriptstyle y:\bottom}
	 \DisplayProof
	  \end{array}
$$
\end{proof}

  \begin{figure*}
$$
    \begin{array}{ccc}	  
    \axc{\pi}
    \noLine
	  \uic{\scriptstyle x\mathcal{R}y}
    \axc{\pi_1}	
    \noLine  
	  \uic{\scriptstyle x=z}
	   \alwaysSingleLine
	   \RightLabel{$\scriptstyle \mathit{mon}$}
    \bic{\scriptstyle z\mathcal{R}y}
    \axc{\pi_2}
    \noLine
	  \uic{\scriptstyle y=u}    
	   \RightLabel{$\scriptstyle \mathit{mon}$}	  
	  \bic{\scriptstyle z\mathcal{R}u}
	  \axc{\pi_3}
	   \noLine
	  \uic{\scriptstyle z=v}
	   \RightLabel{$\scriptstyle \mathit{mon}$}	  
	  \bic{\scriptstyle v\mathcal{R}u}	   
	 \DisplayProof
	    &
	  \rightsquigarrow
	    &
    \axc{\pi}
    \noLine
	  \uic{\scriptstyle x\mathcal{R}y}
    \axc{\pi_1}	
    \noLine  
	  \uic{\scriptstyle x=z}
	   \alwaysSingleLine
	   \RightLabel{$\scriptstyle \mathit{mon}$}
    \bic{\scriptstyle z\mathcal{R}y}
	  \axc{\pi_3}
	   \noLine
	  \uic{\scriptstyle z=v}
	   \RightLabel{$\scriptstyle \mathit{mon}$}	  
	  \bic{\scriptstyle v\mathcal{R}y}    
    \axc{\pi_2}
    \noLine
	  \uic{\scriptstyle y=u}    
	   \RightLabel{$\scriptstyle \mathit{mon}$}	  
	  \bic{\scriptstyle v\mathcal{R}u}	   
	 \DisplayProof
	 \end{array}
$$
\caption{Rule permutation for the ordering of $\mathit{mon}$ applications{\label{fig:mon-ordering}}}
\end{figure*}

 \begin{proof}[Lemma~\ref{lem:pre-normal}] 
We follow the procedure based on proper reductions used in \cite{Vig:Labelled:00} and we only treat the cases $\limpliesI$/$\limpliesE$, $\g I$/$\g E$ and $\forall I$/$\forall E$. The transformations for the detours $\rimpliesI$/$\rimpliesE$ and $\h I$/$\h I$ can be easily inferred from these. Any formula $\varphi$ in a derivation is the root of a tree of rule applications leading back to assumptions. We call \emph{side formulas} of $\varphi$ the formulas in this tree other than $\varphi$. In order to eliminate maximal formulas from a derivation, it suffices to apply the transformations listed below, picking in the set of maximal formulas the formula with the highest grade that has only maximal formulas of lower grade as side formulas, and iterating this process until there are no more maximal formulas in the proof. The process ends because at every step no new maximal formula as large as (or larger than) the eliminated one is introduced.
 $$
	  \begin{array}{cccc}
	  \\
    (i)
    &
	  
	  \axc{\scriptstyle [x:A]}
			\noLine
	  \uic{\scriptstyle \pi_1}
      \noLine
    \uic{\scriptstyle x:B}
			\RightLabel{$\scriptstyle \limplies I$}
		\uic{\scriptstyle x:A\limplies B}
		\axc{\scriptstyle \pi_2}
		  \noLine
		\uic{\scriptstyle x:A}
			\RightLabel{$\scriptstyle \limplies E$}		
	 	\bic{\scriptstyle x:B}
	 \DisplayProof
	 
	    &
	  \rightsquigarrow
	    &
	    
	  \axc{\scriptstyle \pi_2}
			\alwaysNoLine
	 	\uic{\scriptstyle x:A}
	 	\uic{\scriptstyle \pi_1}
	 	\uic{\scriptstyle x:B}
	 \DisplayProof
	  
	  \end{array}
$$
$$	  
	  \begin{array}{cccc}
	  \\
    (ii)
    &
	  
	  \axc{\scriptstyle [x<y]}
			\noLine
	  \uic{\scriptstyle \pi}
      \noLine
    \uic{\scriptstyle y:A}
			\RightLabel{$\scriptstyle \gp I$}
		\uic{\scriptstyle x:\gp A}
		\axc{\scriptstyle x<z}
			\RightLabel{$\scriptstyle \gp E$}		
	 	\bic{\scriptstyle z:A}
	 \DisplayProof
	 
	    &
	  \rightsquigarrow
	    &
	    
	  \axc{\scriptstyle x<z}
			\alwaysNoLine
	 	\uic{\scriptstyle \pi[z/y]}
	 	\uic{\scriptstyle z:A}
	 \DisplayProof

  \end{array}
$$
$$	  
	  \begin{array}{cccc}
	  \\
    (iii)
    &
	  
	  \axc{\scriptstyle \pi}
	    \noLine
	  \uic{\scriptstyle \rho}
	  	\RightLabel{$\scriptstyle \forall I$}
	  \uic{\scriptstyle \forall x.\, \rho}
			\RightLabel{$\scriptstyle \forall E$}	
	  \uic{\scriptstyle \rho[y/x]}
	 \DisplayProof
	 
	    &
	  \rightsquigarrow
	    &
	    
	  \axc{\scriptstyle \pi[y/x]}
			\alwaysNoLine
	 	\uic{\scriptstyle \rho[y/x]}
	 \DisplayProof

  \end{array}
$$
Finally, in Fig.~\ref{fig:mon-ordering} we show how to permute applications of rules in order to get a derivation where, given a sequence of $\mon$ applications, the ones on the same label position occur one immediately below the other. We denote with $\mathcal{R}$ a relational symbol that can stay both for $<$ and for $=$. In the derivation on the left, the first and the third application of $\mathit{mon}$ refer to the same label position and thus are moved one immediately below the other. The derivations obtained in this way will then be further simplified during the normalization process.
\end{proof}

\begin{proof}[Theorem~\ref{th:kl/normal-form}]
First, we observe that by Lemma~\ref{lem:pre-normal} we can obtain a derivation in pre-normal form. Now let us show how to remove redundant formulas.
We know from Lemma~\ref{lem:kl/restrictions} that every application of a falsum-rule has an atomic formula as a conclusion. Thus it is sufficient to consider the following transformations:
 $$
	  \begin{array}{cccc}
	  \\
    (i)
    &
	  \axc{\scriptstyle \Gamma \, \Delta}
	   \noLine
	  \uic{\scriptstyle \pi}
	   \noLine	  
	  \uic{\scriptstyle x:\bottom}
			\RightLabel{$\scriptstyle \bottomE$}	  
	  \uic{\scriptstyle y:\bottom}
			\RightLabel{$\scriptstyle \bottomE$}
	 	\uic{\scriptstyle z:A}
	 \DisplayProof
	 
	    &
	  \rightsquigarrow
	    &

	  \axc{\scriptstyle \Gamma \, \Delta}
	   \noLine
	  \uic{\scriptstyle \pi}
	   \noLine
	  \uic{\scriptstyle x:\bottom}
			\RightLabel{$\scriptstyle \bottomE$}
	 	\uic{\scriptstyle z:A}
	 \DisplayProof
	  \end{array}
$$
where $A$ is $\bottom$ or an atomic formula. Note that if the formula $z:A \limplies \bottom$ is contained in $\Gamma$ and discharged by the second application of $\bottomE$ in the derivation on the left, then the same can be done in the derivation on the right.
$$	  
	  \begin{array}{cccc}
	  \\
    (ii)
    &
	  \axc{\scriptstyle \pi}
	   \noLine	  	  
	  \uic{\scriptstyle x:\bottom}
			\RightLabel{$\scriptstyle \bottomE$}	  
	  \uic{\scriptstyle y:\bottom}
			\RightLabel{$\scriptstyle \mathit{uf}1$}
	 	\uic{\scriptstyle \emptyset}
	 \DisplayProof
	 
	    &
	  \rightsquigarrow
	    &
	  \axc{\scriptstyle \pi}
	   \noLine	  	    
	  \uic{\scriptstyle x:\bottom}
			\RightLabel{$\scriptstyle \mathit{uf}1$}
	 	\uic{\scriptstyle \emptyset}
	 \DisplayProof
	 
	  \end{array}
$$
$$	  
	  \begin{array}{cccc}
	  \\
    (iii)
    &
	  \axc{\scriptstyle \pi}
	   \noLine	  	  
	  \uic{\scriptstyle x:\bottom}
			\RightLabel{$\scriptstyle \mathit{uf}1$}	  
	  \uic{\scriptstyle \emptyset}
			\RightLabel{$\scriptstyle \mathit{uf}2$}
	 	\uic{\scriptstyle y:\bottom}
	 \DisplayProof
	 
	    &
	  \rightsquigarrow
	    &
	  \axc{\scriptstyle \pi}
	   \noLine	  	    
	  \uic{\scriptstyle x:\bottom}
			\RightLabel{$\scriptstyle \bottomE$}
	 	\uic{\scriptstyle y:\bottom}
	 \DisplayProof	    
  
	  \end{array}
$$
$$	  
	  \begin{array}{cccc}
	  \\
    (iv)
    &
	  \axc{\scriptstyle \pi}
	   \noLine	  	  
	  \uic{\scriptstyle \emptyset}
			\RightLabel{$\scriptstyle \mathit{uf}2$}	  
	  \uic{\scriptstyle x:\bottom}
			\RightLabel{$\scriptstyle \mathit{uf}1$}
	 	\uic{\scriptstyle \emptyset}
	 \DisplayProof
	 
	    &
	  \rightsquigarrow
	    &
	  \axc{\scriptstyle \pi}
	   \noLine	  	    
	  \uic{\scriptstyle \emptyset}
	 \DisplayProof	   
	   
	  \end{array}
 $$
For the rule $\mon$, given the ordering of $\mathit{mon}$ applications obtained by permutations defined in Lemma~\ref{lem:pre-normal}, the only case we have to treat is when two applications of $\mon$ working on the same label position of a formula occur consecutively. Then we simply exploit the transitivity of $=$ (obtained by using $\mathit{mon}$). Note that, by Lemma~\ref{lem:kl/restrictions}, in the following reduction $\varphi$ is an atomic formula.
 $$
 \begin{array}{ccc}
\def\defaultHypSeparation{\hskip .00000001in} 
	  \axc{\scriptstyle \pi_1}
	   \noLine	  	  
	  \uic{\scriptstyle \varphi}
	  \axc{\scriptstyle \pi_2}
	   \noLine	  	  	  
	  \uic{\scriptstyle x=y}
			\RightLabel{$\scriptstyle \mon$}
	  \bic{\scriptstyle \varphi[y/x]}
	  \axc{\scriptstyle \pi_3}
	   \noLine	  	  	  
	  \uic{\scriptstyle y=z}
			\RightLabel{$\scriptstyle \mon$}
	 	\bic{\scriptstyle \varphi[z/x]}
	 \DisplayProof
	 
	    &
	  \rightsquigarrow
	    &
\def\defaultHypSeparation{\hskip .00000001in}	    
	  \axc{\scriptstyle \pi_1}
	   \noLine	  	  
	  \uic{\scriptstyle \varphi}
	  \axc{\scriptstyle \pi_2}
	   \noLine	  	  	  
	  \uic{\scriptstyle x=y}
	  \axc{\scriptstyle \pi_3}
	   \noLine	  	  	  
	  \uic{\scriptstyle y=z}
			\RightLabel{$\scriptstyle \mon$}
	  \bic{\scriptstyle x=z}
			\RightLabel{$\scriptstyle \mon$}
	 	\bic{\scriptstyle \varphi[z/x]}
	 \DisplayProof
	   
	  \end{array}
 $$ 
\end{proof}

\begin{proof}[Lemma~\ref{lem:kl/tracks-properties}]
(i) and (ii) follow from the absence of maximal formulas in a normal derivation: in a track $t$, no introduction rule application can precede an application of an elimination rule. In other words, a track in a normal derivation is such that the elimination part (when not empty) starts with a non-atomic formula and consists of some applications of elimination-rules; if the elimination part ends with an atomic formula, then the central part (when not empty) consists of some applications of rules whose conclusion is still an atomic formula; the introduction part (when not empty) starts with an atomic formula and consists of some applications of introduction rules (see Fig~\ref{fig:kl/clessidre}).

(iii) comes from the fact that in a normal derivation a falsum-rule and the $\mon$-rule can be applied only to atomic formulas.

(iv) follows directly from the absence of redundant formulas in a normal derivation (see Theorem~\ref{th:kl/normal-form}).

For (v) and (vi), observe that tracks originating from an application of $\mathit{uf}1$ or $\mathit{uf}2$ start with an atomic formula and thus cannot have an elimination part, while tracks ending in an application of $\mathit{uf}1$ or $\mathit{uf}2$ end with an atomic formula and thus their introduction part must be empty.
\end{proof}

\end{document}